\documentclass[3p,times]{elsarticle}
\usepackage{amsmath}
\usepackage{xcolor}
\usepackage{booktabs}
\usepackage{subcaption}
\usepackage{verbatim}
\usepackage{multirow}
\usepackage{lineno}
\usepackage{url}

\journal{arXiv}
\begin{document}

\begin{frontmatter}
\title {Dual Residual Attention Network for Image Denoising}

\author[mymainaddress]{Wencong Wu}

\author[mymainaddress]{Shijie Liu}

\author[mymainaddress]{Yi Zhou}

\author[mymainaddress]{Yungang Zhang\corref{mycorrespondingauthor}}
\ead{yungang.zhang@ynnu.edu.cn}

\author[mymainaddress]{Yu Xiang\corref{mycorrespondingauthor}}
\ead{xiangyu@ynnu.edu.cn}

\cortext[mycorrespondingauthor]{Corresponding author}

\address[mymainaddress]{School of Information Science and Technology, Yunnan Normal University, Kunming 650500, Yunnan Province, China}

\begin{abstract}
In image denoising, deep convolutional neural networks (CNNs) can obtain favorable performance on removing spatially invariant noise. However, many of these networks cannot perform well on removing the real noise (i.e. spatially variant noise) generated during image acquisition or transmission, which severely sets back their application in practical image denoising tasks. Instead of continuously increasing the network depth, many researchers have revealed that expanding the width of networks can also be a useful way to improve model performance. It also has been verified that feature filtering can promote the learning ability of the models. Therefore, in this paper, we propose a novel Dual-branch Residual Attention Network (DRANet) for image denoising, which has both the merits of a wide model architecture and attention-guided feature learning. The proposed DRANet includes two different parallel branches, which can capture complementary features to enhance the learning ability of the model. We designed a new residual attention block (RAB) and a novel hybrid dilated residual attention block (HDRAB) for the upper and the lower branches, respectively. The RAB and HDRAB can capture rich local features through multiple skip connections between different convolutional layers, and the unimportant features are dropped by the residual attention modules. Meanwhile, the long skip connections in each branch, and the global feature fusion between the two parallel branches can capture the global features as well. Moreover, the proposed DRANet uses downsampling operations and dilated convolutions to increase the size of the receptive field, which can enable DRANet to capture more image context information. Extensive experiments demonstrate that compared with other state-of-the-art denoising methods, our DRANet can produce competitive denoising performance both on synthetic and real-world noise removal. The code for DRANet is accessible at https://github.com/WenCongWu/DRANet.
\end{abstract}


\end{frontmatter}

\section{Introduction}
Image denoising is a classical inverse problem in low-level vision and image processing. Its goal is to restore a high-quality image $x$ from a noisy image $y$ containing noise $n$, which can be expressed as $x = y - n$. During the past decades, several denoising models have been developed. For instance, the famous BM3D algorithm \cite{Dabov2007} uses the non-local self-similarity and sparse representation to improve denoising performance. The weighted nuclear norm minimization (WNNM) \cite{Gu2014} utilizes prior knowledge and the low rank theory to promote the denoising effect. Xu et al. \cite{Xu2017} designed multi-channel weighted nuclear norm minimization (MCWNNM) for practical image denoising. Later, the same authors proposed the trilateral weighted sparse coding (TWSC) to achieve excellent practical image denoising performance by using noise modeling and image priors \cite{Xu2018}. Although these traditional denoising models can obtain good denoising results, they are difficult to be applied in practice due to their long inference time caused by their complex optimization algorithms, and the unstable denoising performance due to the large number of hyper-parameters.

With the advent of DNNs, many DNN-based denoising models have successfully addressed the issues of the traditional denoising models. The DNN-based models have fewer hyperparameters and shorter inference time to achieve excellent denoising performance. For example, the famous denoising convolutional neural network (DnCNN) \cite{Zhang2017} achieved remarkable denoising performance compared to the traditional methods. The FFDNet \cite{Zhang2018} later introduces a fast and flexible image denoising solution. On the basis of the FFDNet, Wu et al. \cite{Wu2023} proposed a more flexible and efficient U-shaped network (FEUNet). The FFDNet and FEUNet have great flexibility for spatially variant or invariant additive Gaussian white noise (AWGN).

Although a deeper network may have more powerful learning ability, some negative effects may also be brought. As the network gets deeper, the number of parameters and denoising inference time will also be increased correspondingly, which is difficult to apply to real-world denoising scenes. To address the issue, many researchers have tried to improve the representation ability of the denoising networks by widening the network structures, and promising denoising results were reported \cite{Pan2018, Tian2020, Tian2021}.

Many attention-based denoising models have also been presented. For example, Tian et al. \cite{TianX2020} developed an attention-guided denoising network (ADNet). Anwar et al. \cite{Anwar2019} proposed a single-stage RIDNet model for real image denoising. Although these methods achieve excellent performance, their feature learning ability can be further enhanced. Skip connection \cite{He2016} has been verified as a useful tool to improve feature learning ability. Peng et al. \cite{Peng2019} proposed a DSNet model with symmetric skip connection for image denoising. Zhang et al. \cite{Zhang2021} presented a RDN model with dense connections for image denoising. These attention-guided denoising models have proven that feature filtering is an effective tool for image denoising.

Inspired by the success of dual CNN structure and attention mechanism in image denoising, in this paper, we develop a new dual residual attention network (DRANet) for image noise removal. The DRANet consists of two different sub-networks, which are designed to extract complementary features to enhance model performance. The upper sub-network mainly includes 5 residual attention blocks (RABs) with multiple skip connections, and the lower sub-network mainly contains 5 hybrid dilated residual attention blocks (HDRABs), where skip connections are also applied. The designed RAB and HDRAB can extract rich local features and implement feature filtering. Moreover, the long skip connections in each branch and the concatenation between two sub-networks can exploit global features. Overall, the contributions of this paper are summarized as follows:

(1) We design a novel dual residual attention network (DRANet) for image blind denoising, which is effective for both synthetic noise and real-world noise. Moreover, the DRANet has a simpler architecture and fewer network parameters compared with many state-of-the-art blind denoising models.

(2) We propose the residual attention block (RAB) and the hybrid dilated residual attention block (HDRAB) for the two branches of our model, respectively. This gives the model the ability to capture complementary features. More importantly, the designed attention blocks can implement appropriate feature filtering, and ensure the whole network to extract both local and global features well.

(3) Extensive experiments have verified that the DRANet can obtain competitive denoising results compared to many other state-of-the-art denoising methods.

The remainder of this paper is organized as follows. In Section \ref{Related_work}, we review the related denoising models. The proposed DRANet is introduced in Section \ref{Proposed_model}. In Section \ref{Experiment}, we present the experimental results. This paper is summarized in Section \ref{Conclusion}.

\section{Related work}\label{Related_work}

\subsection{Blind denoising}
Many existing denoising models have to take manually preset parameters (i.e. noise level) as their input to obtain better denoising performance, such as the FFDNet \cite{Zhang2018}, FEUNet \cite{Wu2023}, CSANN \cite{Wang2021}, and DRUNet \cite{Zhang2022}, which is extremely inflexible and therefore restricts the application of these models in real-world denoising tasks. To address this problem, many blind denoising models have been developed, such as the CBDNet \cite{Guo2019}, BUIFD \cite{Helou2020}, VDN \cite{Yue2019}, DUBD \cite{Soh2020}, AINDNet \cite{Kim2020}, DCBDNet \cite{WuS2023}, and DCANet \cite{WuG2023}. All of these models utilize a noise estimator to predict noise level in a noisy image, which can significantly increase the flexibility of the model. It is apparent that the noise estimators in these models play a crucial role. However, it is still very challenging to obtain an accurate noise estimator. Meanwhile, an additional noise estimator may increase the complexity of a denoising model.

Besides the aforementioned blind denoising models equipped with a noise estimator, other blind denoising techniques have also been proposed. A denoising CNN (DnCNN) \cite{Zhang2017} was designed and can be used for image blind denoising, the batch normalization (BN) \cite{Ioffe2015} and residual learning \cite{He2016} were adopted to promote the network performance, and accelerate the network training. An image restoration CNN (IRCNN) was presented in \cite{ZhangZGZ2017}, where the BN and residual learning were also applied to improve network performance. Anwar et al. \cite{Anwar2019} proposed a RIDNet model for real image blind denoising, where the enhancement attention module (EAM) with an enhanced residual block and a feature attention module was designed as its backbone. Li et al. \cite{Li2022} proposed an AirNet model to achieve multiple low-quality image restoration tasks by using only a single network. A spatial-adaptive denoising network (SADNet) was designed in \cite{Chang2020} for image blind denoising, where the deformable convolution \cite{Dai2017, Zhu2019} was used to adjust the convolution kernel size to fit the spatial support area of the network. Yue et al. \cite{Yue2020} presented a dual adversarial network (DANet) for image denoising. Soh et al. \cite{Soh2022} developed a universal variational approach for image blind restoration tasks, which is named as variational deep image restoration (VDIR), the model decomposes the restoration problems into easy-to-solve sub-problems based on variational approximation.

\subsection{Residual learning and dilated convolution}
Residual learning \cite{He2016} was initially developed for addressing the problem of performance degradation in image recognition tasks as the networks become deeper, and residual learning can also effectively prevent the vanishing or exploding of gradients. Later, many versions of the residual network (ResNet) \cite{Zhang2021, Kim2016, ZhangT2018, Kokkinos2019} were developed for image restoration tasks. Due to the use of residual learning, the increase in network depth has a relatively small impact on their performance, and a larger receptive field can be produced to further promote the extraction of more useful image information. However, deeper networks often bring more network parameters and longer inference time, which makes it difficult for the very deep models to be deployed on the devices with limited storage space, or to be applied in scenarios with high real-time requirements.

To overcome this drawback, the dilated convolution \cite{Yu2015} was designed to expand the receptive field of the nework without increasing the network complexity. Tian et al. \cite{Tian2020} developed a batch-renormalization denoising network (BRDNet), where the dilated convolution with a fixed dilated rate is adopted to enlarge the receptive field. However, the dilated convolution tends to cause the gridding phenomenon. Therefore, the hybrid dilated filters \cite{Yu2017, Wang2018} were developed to address this issue. In the IRCNN \cite{ZhangZGZ2017}, the hybrid dilated convolution was used to increase its receptive field and promote its restoration performance. Later, Peng et al. \cite{Peng2019} designed a dilated residual network with symmetric skip connection (DSNet) for image denoising, where the dilated convolutions with different dilated rates were utilized to enhance network performance, and symmetric skip connection was employed to obtain more image details and avoid the gradient vanishing or exploding.

\subsection{Attention mechanism}
To extract and select proper features is crucial for image processing \cite{Du2019, Li2021, Liang2021}. However, obtaining useful features is not easy for images with complex backgrounds \cite{Li2019}. Therefore, the attention mechanisms \cite{Hu2018, Woo2018} were developed to extract informative features better, and they have been widely utilized in many computer vision fields \cite{ZhangL2018, Dai2019, Lyn2020}, including image denoising.

An attention-guided denoising network (ADNet) was presented in \cite{TianX2020}, where the noise information is captured by the attention block. A single-stage RIDNet model was designed for real image blind denoising \cite{Anwar2019}, in which the weights of informative features are emphasized by a feature attention module. A residual dilated attention network (RDAN) contains multiple residual dilated attention blocks (RDAB) and residual conv attention blocks (RCAB) was proposed in \cite{Hou2019}, which includes non-local and local operations to promote overall feature extraction. In \cite{Zhang2019}, the residual non-local attention networks (RNAN) were designed for image restoration, where the residual local and non-local attention blocks were applied to exploit the long-range dependencies between pixels. The DeamNet was proposed in \cite{Ren2021} for image blind denoising, in which a dual element-wise attention mechanism (DEAM) was designed to enhance its expressive ability. Furthermore, some denoising models such as the CSANN \cite{Wang2021}, CycleISP \cite{Zamir2020}, and DCANet \cite{WuG2023}, employ channel attention \cite{Hu2018} and spatial attention \cite{Woo2018} mechanisms to learn inter-channel and inter-spatial relationships between convolutional features, the denoising performance is therefore improved.

\subsection{Dual denoising network}
Some researchers have verified that using the dual network structure can also achieve promising image denoising performance. A DualCNN \cite{Pan2018, Pan2022} denoising network was proposed for low-level vision tasks, where two parallel branches are utilized to extract image structures and details, respectively. Tian et al. \cite{Tian2020} designed a batch-renormalization denoising network (BRDNet), which contains two different sub-networks, and batch-renormalization \cite{Ioffe2017} and residual learning were adopted in the BRDNet model to improve its denoising ability. Later, a dual denoising network (DudeNet) \cite{Tian2021} was introduced, which fuses local and global features to restore image details better. Wu et al. \cite{WuS2023} developed a dual convolutional blind denoising network with multiple skip connections (DCBDNet), and the DCBDNet is composed of two different sub-networks with a similar receptive field, which not only saves the computational cost, but also promotes its denoising performance. They further proposed a novel dual convolutional neural network with attention (DCANet) \cite{WuG2023} for image blind denoising, which is the first work exploring the combination of dual CNN and attention mechanism.

\section{The proposed model}\label{Proposed_model}
In this section, we first introduce the whole structure of the DRANet model, followed by more details of the Residual Attention Block (RAB) and Hybrid Dilated Residual Attention Block (HDRAB).

\subsection{Network architecture}
\begin{figure*}[htbp]
	\begin{center}
		\includegraphics[width=\textwidth]{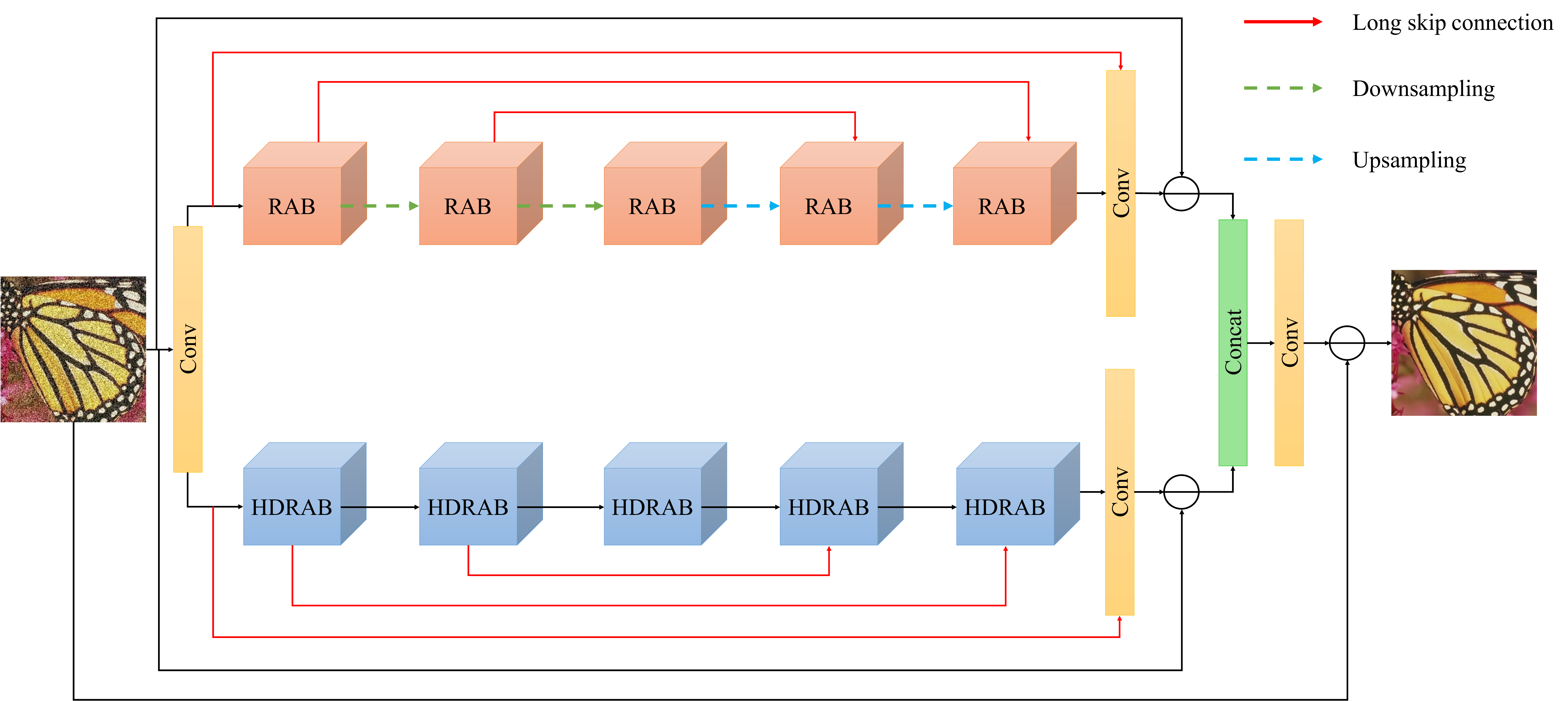}
		\caption{The network architecture of the DRANet for image denoising.}
		\label{fig:DRANet}
	\end{center}
\end{figure*}

The structure of the DRANet model is shown in Fig. \ref{fig:DRANet}. It can be found that the model mainly contains two different sub-networks. The upper sub-network contains five RABs and a convolutional layer, the $2 \times 2$ strided convolutions and $2 \times 2$ transposed convolutions are employed for image downsampling and upsampling, respectively. By implementing the downsamplings and upsamplings, the multi-scale features can be extracted, and the receptive field can be increased. The lower sub-network is composed of five HDRABs and a convolutional layer. The symbol `$\ominus$' and `Concat' denote residual learning and concatenation operation, respectively. Long skip connections are used between the RABs and between the HDRABs. The features learned by the two branches are fused by using the concatenation operation, therefore the global image features can be captured. The long skip connections and residual learning \cite{He2016} in the model can not only accelerate network training, but also improve its performance. The size of all convolutional filter kernels in the model is set to $3 \times 3 \times 128$.

\subsection{Residual attention block}
\begin{figure*}[htbp]
	\begin{center}
		\includegraphics[width=\textwidth]{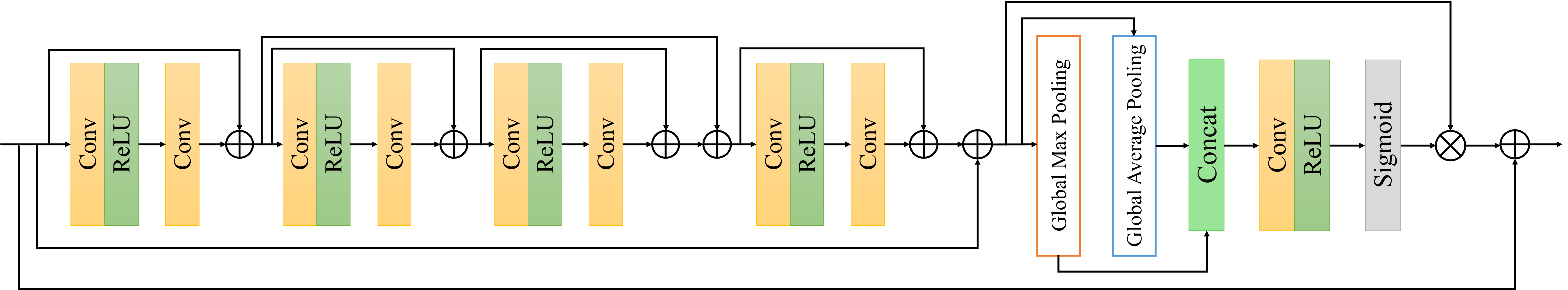}
		\caption{The architecture of the residual attention block (RAB).}
		\label{fig:RAB}
	\end{center}
\end{figure*}

The designed residual attention block (RAB) is displayed in Fig. \ref{fig:RAB}. The RAB consists of the residual block and spatial attention module (SAM) \cite{Woo2018}. The residual block includes multiple standard convolutions (Conv) and rectified linear units (ReLU) \cite{Krizhevsky2012}. The rich local features can be extracted via multiple skip connections between convolutional layers, where the symbol `$\oplus$' in Fig. \ref{fig:RAB} denotes the element-wise addition. Since convolutional layers can only extract the local information and cannot exploit the non-local information, which may lead to degraded denoising performance, the attention mechanism is applied to capture and learn the global context information. Compared with the existing residual attention block \cite{Hou2019, Zhang2019}, our RAB owns more skip connections, which can extract and fuse features between different convolutional layers to further improve the denoising performance.

The SAM module is used to learn the inter-spatial relationships between the convolutional features, where the symbol $\otimes$ represents the element-wise product. The SAM contains the Global Max Pooling (GMP), Global Average Pooling (GAP), Conv, ReLU, and Sigmoid \cite{Han1995}. The GAP and GMP are used in our RAB to represent the statistics of the whole image, so that the features with more useful information are focused and the uninformative features are filtered. It should be noted that by using residual learning, very deep networks can be used to promote denoising performance, however the network complexity will also be greatly increased. To balance the network complexity and denoising performance, five RABs are used in our model.

\subsection{Hybrid dilated residual attention block}
\begin{figure*}[htbp]
	\begin{center}
		\includegraphics[width=\textwidth]{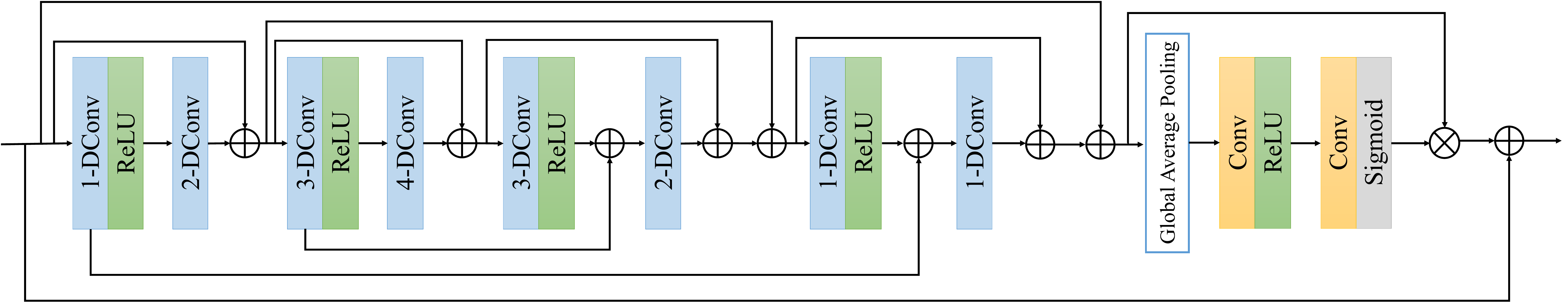}
		\caption{The architecture of the hybrid dilated residual attention block (HDRAB).}
		\label{fig:HDRAB}
	\end{center}
\end{figure*}

We designed the hybrid dilated residual attention block (HDRAB) for the lower branch of our model, which is displayed in Fig. \ref{fig:HDRAB}. The HDRAB is composed of the hybrid dilated residual block and channel attention module (CAM) \cite{Hu2018}. The hybrid dilated residual block contains multiple hybrid dilated convolutions ($s$-DConv in Fig. \ref{fig:HDRAB}) \cite{Yu2015} and ReLU, which can capture the local features via multiple skip connections between the dilated convolutional layers, where the `$s$' represents the dilated rate, whose value ranges from 1 to 4, and the symbol `$\oplus$' represents the element-wise addition. The dilated convolution can expand the size of the receptive field to capture more image information, and the hybrid dilated convolution \cite{Wang2018} was used for eliminating the possible gridding phenomenon. As far as we know, there is no similar block structure to our HDRAB.

The CAM module consists of the GAP, Conv, ReLU, and Sigmoid. The CAM is applied to exploit the inter-channel relationships between the convolutional features, where the symbol `$\otimes$' in Fig. \ref{fig:HDRAB} represents the element-wise product. In order to obtain an appropriate balance between the complexity and performance of the network, we equip the lower sub-network with five HDRABs.

\subsection{Loss function}
To train the Gaussian denoising DRANet model, we adopt the mean squared error (MSE) to optimize our model. The MSE calculates the difference between the predicted image $\hat{x}_j$ and the ground-truth image $x_j$, and is defined as:
\begin{equation}
\begin{aligned}
\mathcal{L} &= \frac{1}{2N} \sum_{j=1}^N \left\|f(y_j, \theta)- x_j\right\|^2\\
            &= \frac{1}{2N} \sum_{j=1}^N \left\|\hat{x}_j - x_j\right\|^2,
\end{aligned}
\end{equation}
\noindent where $y_i$ and $\theta$ are noisy images and the trainable parameters of DRANet, respectively. $N$ is the number of clean-noisy image patches used for training.

To train the DRANet model for real noise removal, using the MSE as the loss function can lead to blurring effects, causing the loss of image details and high-frequency textures in the denoised image. Therefore, we adopt Charbonnier loss \cite{Lai2017} as the reconstruction loss function, and the edge loss \cite{Jiang2020} is used to prevent high-frequency information loss between the denoised image $\hat{x}$ and the ground-truth image $x$. The overall loss function is designed as:

\begin{equation}
\mathcal{L} = \mathcal{L}_{char}(\hat{x}, x) + \lambda_{edge}\mathcal{L}_{edge}(\hat{x}, x),
\label{eq.2}
\end{equation}
\noindent where we empirically set $\lambda_{edge}$ to 0.1. $\mathcal{L}_{char}$ is the Charbonnier loss, represented as:

\begin{equation}
\mathcal{L}_{char} = \sqrt{\left\|\hat{x} - x\right\|^2 + \epsilon^2},
\end{equation}
\noindent where the constant $\epsilon$ is set as $10^{-3}$, and $\mathcal{L}_{edge}$ is the edge loss, which is denoted as:
\begin{equation}
\mathcal{L}_{edge} = \sqrt{\left\|\bigtriangleup{(\hat{x})} - \bigtriangleup{(x)}\right\|^2 + \epsilon^2},
\end{equation}
\noindent where the $\bigtriangleup$ denotes the Laplacian operator \cite{Kamgar1999}.

\section{Experiments and results}\label{Experiment}
\subsection{Datasets}
Seven public image datasets were used in our experiments for performance evaluation. For the synthetic noise removal, we used the Flick2K dataset \cite{Lim2017} to train the DRANet model, the dataset consists of 2650 high-resolution (HR) color images. To avoid the waste of computing resources and assure adequate training of the model, we carefully calculated the receptive field size of the DRANet model, then randomly cropped the HR images into image patches of the size $128 \times 128$ for color image denoising. The cropped color image patches were grayscaled for denoising performance evaluation on grayscale images. To generate the synthetic noisy image patches, we randomly add the AWGN with noise levels in the range of $[0, 50]$ to the clean image patches. Besides, we used rotation and flipping operations for data augmentation. To evaluate our DRANet model for the AWGN removal, we used five public datasets as the test sets, which are the Set12 \cite{Roth2005}, BSD68 \cite{Roth2005}, CBSD68 \cite{Roth2005}, Kodak24 \cite{Kodak24}, and McMaster \cite{Zhang2011}.

For real noise elimination evaluation, we chose the SIDD medium dataset \cite{Abdelhamed2018} as the training set for the proposed DRANet, the dataset is composed of 320 pairs of HR noisy images and their near noise-free images. We arbitrarily cropped these HR images into the image patches of size $128 \times 128$, and utilized rotation and flipping operations to augment the training samples. To test our DRANet model for real image denoising, we select the SIDD validation set \cite{Abdelhamed2018} and DND sRGB dataset \cite{Plotz2017}.

\subsection{Experimental settings}
The PyTorch framework and an NVIDIA GeForce RTX 3080Ti GPU are used for model training. For the AWGN removal evaluation, we trained the DRANet models for grayscale and color images separately. It costed about 65 and 99 hours to train the models, respectively. About 190 hours were costed on the training of the DRANet model for real noise.

The parameters of the DRANet models were optimized by the Adam optimizer \cite{Kingma2014}. For the synthetic noise removal models, we trained the DRANet models for a total of $6\times10^{5}$ iterations, the initial learning rate ($lr$) is set to $1\times10^{-4}$, and then started to be halved after every $1\times10^{5}$ iterations. For the real noise removal model, we applied 120 epochs to train the DRANet, and the initial $lr$ is set to $2\times10^{-4}$, which is gradually declined to $1\times10^{-6}$ by using cosine annealing strategy \cite{Loshchilov2017}. For both synthetic and real noise models, in every training batch, 8 image patches of size $128 \times 128$ were fed into the model. For other hyper-parameters of the Adam algorithm, the default settings were utilized.

\subsection{Ablation study}
To verify the effectiveness of the proposed network architecture, we trained five different networks in our ablation study, namely the upper sub-network only, lower sub-network only, DRANet model without long skip connection, DRANet model without residual learning, and the whole DRANet model. We selected the BSD68 dataset \cite{Roth2005} for our ablation experiment, and set the noise level to 30. Table \ref{tab:BSD68_val} shows the average PSNR and the SSIM \cite{Wang2004} of the five different models.

\begin{table*}[htbp]
\centering
\caption{Quantitative results (PSNR) for the AWGN removal evaluation. The best result are bolded.}
\label{tab:BSD68_val}
\begin{tabular}{lll}
\hline
Models & PSNR & SSIM\\
\hline
Upper sub-network only& 27.81 & 0.781\\
\hline
Lower sub-network only& 28.43 & 0.807\\
\hline
DRANet without long skip connection & 28.55 & $\mathbf{0.808}$\\
\hline
DRANet without residual learning & 28.54 & 0.807\\
\hline
DRANet (whole) & $\mathbf{28.56}$ & $\mathbf{0.808}$\\
\hline
\end{tabular}
\end{table*}

In Table \ref{tab:BSD68_val}, one can find that the whole DRANet model obtains the best denoising performance than the other four models. On the average PSNR values, the noise removal performance of the whole DRANet model exceeds that of the upper and lower sub-networks by 0.75 dB and 0.13 dB, respectively. Although the denoising performance of the whole DRANet model is just slightly improved by the long skip connection and residual learning, the long skip connection and residual learning can accelerate network training and convergence. In summary, the results of our ablation experiment show that our proposed DRANet model is effective and efficient.

\subsection{The additive Gaussian noise removal evaluation}
In the subsection, we present the denoising results of our DRANet model on the synthetic noisy images. The BM3D \cite{Dabov2007}, TNRD \cite{Chen2017}, DnCNN \cite{Zhang2017}, IRCNN \cite{ZhangZGZ2017}, FFDNet \cite{Zhang2018}, BUIFD \cite{Helou2020}, BRDNet \cite{Tian2020}, DudeNet \cite{Tian2021}, ADNet \cite{TianX2020}, DSNetB \cite{Peng2019}, AINDNet \cite{Kim2020}, RIDNet \cite{Anwar2019}, AirNet \cite{Li2022}, and VDN \cite{Yue2019} were utilized for comparison.

Table \ref{tab:Set12_PSNR} shows the PSNR values of different denoising methods, where we compared the performances of these denoising methods at three noise levels on the Set12 dataset \cite{Roth2005}. It can be seen that the DRANet achieved competitive average performance at noise level 15, and produced the top average PSNR results at noise levels 25 and 50.

\begin{table*}[htbp]\scriptsize
\centering
\caption{Results (PSNR) of the AWGN removal evaluation on grayscale images. The top two results are emphasized in red and blue, respectively.}
\label{tab:Set12_PSNR}
\begin{tabular}{|c|c|c|c|c|c|c|c|c|c|c|c|c|c|c|}
\hline
Noise levels & Methods & C.man & House & Peppers & Starfish &  Monar. &  Airpl. & Parrot &  Lena &  Barbara &  Boat &  Man & Couple & Average\\
\hline
\hline
\multirow{10}*{$\sigma$=15} & BM3D \cite{Dabov2007} & 31.91 & 34.93 & 32.69 & 31.14	& 31.85	& 31.07	& 31.37	& 34.26	& \textcolor{red}{33.10}	& 32.13	& 31.92	& 31.10	& 32.37\\
\cline{2-15}
    & TNRD \cite{Chen2017} & 32.19	& 34.53	& 33.04	& 31.75	& 32.56	& 31.46	& 31.63	& 34.24	& 32.13	& 32.14	& 32.23	& 32.11	& 32.50\\
\cline{2-15}
    & DnCNN-S \cite{Zhang2017} & 32.61 & 34.97 & 33.30 & 32.20 & 33.09 & 31.70 & 31.83 & 34.62 & 32.64 & 32.42 & 32.46 & 32.47 & 32.86\\
\cline{2-15}
    & BUIFD \cite{Helou2020} & 31.74	& 34.78	& 32.80	& 31.92	& 32.77	& 31.34 & 31.39	& 34.38	& 31.68	& 32.18	& 32.25	& 32.22	& 32.46\\
\cline{2-15}
    & IRCNN \cite{ZhangZGZ2017} & 32.55 & 34.89 & 33.31	& 32.02	& 32.82	& 31.70	& 31.84	& 34.53	& 32.43	& 32.34	& 32.40	& 32.40	& 32.77\\
\cline{2-15}
    & FFDNet \cite{Zhang2018} & 32.43	& 35.07	& 33.25	& 31.99	& 32.66	& 31.57	& 31.81	& 34.62	& 32.54	& 32.38	& 32.41	& 32.46	& 32.77\\
\cline{2-15}
    & ADNet \cite{TianX2020} & \textcolor{red}{32.81} & 35.22 & \textcolor{red}{33.49} & 32.17 & 33.17 & \textcolor{red}{31.86} & \textcolor{blue}{31.96} & 34.71 & 32.80 & \textcolor{blue}{32.57} & \textcolor{blue}{32.47} & 32.58 & 32.98\\
\cline{2-15}
    & BRDNet \cite{Tian2020} & \textcolor{blue}{32.80} & \textcolor{blue}{35.27} & \textcolor{blue}{33.47} & \textcolor{blue}{32.24} & \textcolor{red}{33.35} & \textcolor{blue}{31.85} & \textcolor{red}{32.00} & \textcolor{blue}{34.75} & \textcolor{blue}{32.93} & 32.55 & \textcolor{red}{32.50} & \textcolor{blue}{32.62} & \textcolor{red}{33.03} \\
\cline{2-15}
    & DudeNet \cite{Tian2021} & 32.71 & 35.13 & 33.38 & \textcolor{red}{32.29} & 33.28 & 31.78 & 31.93 & 34.66 & 32.73 & 32.46 & 32.46 & 32.49 & 32.94\\
\cline{2-15}
    & DRANet & 32.57	& \textcolor{red}{35.42} & 33.32	& 32.19	& \textcolor{blue}{33.31} & 31.78	& 31.95	& \textcolor{red}{34.81}	& 32.91	& \textcolor{red}{32.60}	& \textcolor{red}{32.50}	& \textcolor{red}{32.65}	& \textcolor{blue}{33.00}\\
\hline
\hline
\multirow{10}*{$\sigma$=25} & BM3D \cite{Dabov2007} & 29.45 & 32.85 & 30.16 & 28.56 & 29.25 & 28.42 & 28.93 & 32.07 & \textcolor{red}{30.71} & 29.90 & 29.61 & 29.71 & 29.97 \\
\cline{2-15}
    & TNRD \cite{Chen2017} & 29.72 & 32.53 & 30.57 & 29.02 & 29.85 & 28.88 & 29.18 & 32.00 & 29.41 & 29.91 & 29.87 & 29.71 & 30.06\\
\cline{2-15}
    & DnCNN-S \cite{Zhang2017} & 30.18 & 33.06 & 30.87 & 29.41 & 30.28 & 29.13 & 29.43 & 32.44 & 30.00 & 30.21 & 30.10 & 30.12 & 30.43 \\
\cline{2-15}
    & BUIFD \cite{Helou2020} & 29.42	& 33.03	& 30.48	& 29.21	& 30.20	& 28.99	& 28.94	& 32.20	& 29.18	& 29.97	& 29.88	& 29.90	& 30.12\\
\cline{2-15}
    & IRCNN \cite{ZhangZGZ2017} & 30.08 & 33.06 & 30.88 & 29.27 & 30.09 & 29.12 & 29.47 & 32.43 & 29.92 & 30.17 & 30.04 & 30.08 & 30.38 \\
\cline{2-15}
    & FFDNet \cite{Zhang2018} & 30.10 & 33.28 & 30.93 & 29.32 & 30.08 & 29.04 & 29.44 & 32.57 & 30.01 & 30.25 & 30.11 & 30.20 & 30.44\\
\cline{2-15}
    & ADNet \cite{TianX2020} & \textcolor{blue}{30.34} & \textcolor{blue}{33.41} & \textcolor{red}{31.14} & 29.41 & 30.39 & 29.17 & 29.49 & 32.61 & 30.25 & \textcolor{blue}{30.37} & 30.08 & 30.24 & 30.58\\
\cline{2-15}
    & BRDNet \cite{Tian2020} & \textcolor{red}{31.39} & \textcolor{blue}{33.41} & \textcolor{blue}{31.04} & 29.46 & \textcolor{blue}{30.50} & \textcolor{blue}{29.20} & \textcolor{blue}{29.55} & \textcolor{blue}{32.65} & 30.34 & 30.33 & \textcolor{blue}{30.14} & \textcolor{blue}{30.28} & \textcolor{blue}{30.61}\\
\cline{2-15}
    & DudeNet \cite{Tian2021} & 30.23 & 33.24 & 30.98 & \textcolor{red}{29.53} & 30.44 & 29.14 & 29.48 & 32.52 & 30.15 & 30.24 & 30.08 & 30.15 & 30.52\\
\cline{2-15}
    & DRANet & 30.30	& \textcolor{red}{33.62}	& 30.98	& \textcolor{blue}{29.52}	& \textcolor{red}{30.61}	& \textcolor{red}{29.21}	& \textcolor{red}{29.56}	& \textcolor{red}{32.83}	& \textcolor{blue}{30.54}	& \textcolor{red}{30.49}	& \textcolor{red}{30.19}	& \textcolor{red}{30.40}	& \textcolor{red}{30.69}\\
\hline
\hline
\multirow{10}*{$\sigma$=50} & BM3D \cite{Dabov2007} & 26.13 & 29.69 & 26.68 & 25.04 & 25.82 & 25.10 & 25.90 & 29.05 & \textcolor{blue}{27.22} & 26.78 & 26.81 & 26.46 & 26.72\\
\cline{2-15}
    & TNRD \cite{Chen2017} & 	26.62 & 29.48 & 27.10 & 25.42 & 26.31 & 25.59 & 26.16 & 28.93 & 25.70 & 26.94 & 26.98 & 26.50 & 26.81\\
\cline{2-15}
    & DnCNN-S \cite{Zhang2017} & 27.03 & 30.00 & 27.32 & 25.70 & 26.78 & 25.87 & 26.48 & 29.39 & 26.22 & 27.20 & 27.24 & 26.90 & 27.18\\
\cline{2-15}
    & BUIFD \cite{Helou2020} & 25.44	& 29.76	& 26.50	& 24.87	& 26.49	& 25.34	& 25.07	& 28.81	& 25.49	& 26.59	& 26.87	& 26.34	& 26.46\\
\cline{2-15}
    & IRCNN \cite{ZhangZGZ2017} & 26.88 & 29.96 & 27.33 & 25.57 & 26.61 & 25.89 & 26.55 & 29.40 & 26.24 & 27.17 & 27.17 & 26.88 & 27.14 \\
\cline{2-15}
    & FFDNet \cite{Zhang2018} & 27.05 & 30.37 & 27.54 & 25.75 & 26.81 & 25.89 & 26.57 & \textcolor{blue}{29.66} & 26.45 & 27.33 & \textcolor{blue}{27.29} & 27.08 & 27.32 \\
\cline{2-15}
    & ADNet \cite{TianX2020} & 27.31 & \textcolor{blue}{30.59} & \textcolor{red}{27.69} & 25.70 & 26.90 & 25.88 & 26.56 & 29.59 & 26.64 & 27.35 & 27.17 & 27.07 & 27.37\\
\cline{2-15}
    & BRDNet \cite{Tian2020} & \textcolor{blue}{27.44} & 30.53 & \textcolor{blue}{27.67} & 25.77 & \textcolor{blue}{26.97} & \textcolor{blue}{25.93} & \textcolor{blue}{26.66} & 25.93 & 26.66 & \textcolor{blue}{27.38} & 27.27 & \textcolor{blue}{27.17} & \textcolor{blue}{27.45}\\
\cline{2-15}
    & DudeNet \cite{Tian2021} & 27.22 & 30.27 & 27.51 & \textcolor{blue}{25.88} & 26.93 & 25.88 & 26.50 & 29.45 & 26.49 & 27.26 & 27.19 & 26.97 & 27.30\\
\cline{2-15}
    & DRANet & \textcolor{red}{27.58} & \textcolor{red}{30.89}	& 27.62	& \textcolor{red}{25.99}	& \textcolor{red}{27.13}	& \textcolor{red}{26.03}	& \textcolor{red}{26.67}	& \textcolor{red}{29.95}	& \textcolor{red}{27.34}	& \textcolor{red}{27.57}	& \textcolor{red}{27.34}	& \textcolor{red}{27.36}	& \textcolor{red}{27.62}\\
\hline
\end{tabular}
\end{table*}

The averaged SSIM results on the Set12 dataset of the compared methods are reported in Table \ref{tab:Set12_SSIM}, where three noise levels were also used for evaluation. We can find that the average SSIM values of the DRANet, BRDNet, and RIDNet are equal at the noise level of 15. The denoising performance of the DRANet outperforms all other methods at noise levels 25 and 50.

\begin{table}[htbp]
\centering
\caption{Results (SSIM) for the AWGN removal evaluation on grayscale images. The top two results are highlighted in red and blue, respectively.}
\label{tab:Set12_SSIM}
\begin{tabular}{cccc}
\hline
Noise level & $\sigma$=15 & $\sigma$=25 & $\sigma$=50 \\
\hline
BM3D \cite{Dabov2007} & 0.896 & 0.851 & 0.766\\
\hline
TNRD \cite{Chen2017} & 0.896 &  0.851 & 0.768 \\
\hline
DnCNN-S \cite{Zhang2017} & 0.903 & 0.862 & 0.783 \\
\hline
BUIFD \cite{Helou2020} &  0.899 & 0.855 & 0.755 \\
\hline
IRCNN  \cite{ZhangZGZ2017} & 0.901 & 0.860 & 0.780 \\
\hline
FFDNet \cite{Zhang2018} & 0.903 & 0.864 & 0.791 \\
\hline
BRDNet \cite{Tian2020} & \textcolor{red}{0.906} & 0.866 & \textcolor{blue}{0.794} \\
\hline
ADNet \cite{TianX2020} & \textcolor{blue}{0.905} & 0.865 & 0.791 \\
\hline
RIDNet \cite{Anwar2019} & \textcolor{red}{0.906} & \textcolor{blue}{0.867} & 0.793 \\
\hline
DRANet & \textcolor{red}{0.906} & \textcolor{red}{0.868} & \textcolor{red}{0.800} \\
\hline
\end{tabular}
\end{table}

We also evaluated different denoising models on the BSD68 dataset \cite{Roth2005}. The average PSNR and SSIM values are listed in Table \ref{tab:BSD68}. One can find that the average PSNR result of DRANet is slightly lower than that of the RIDNet by 0.02 dB at noise level 15, however the DRANet obtained leading performances at noise levels 25 and 50.

\begin{table}[htbp]
\centering
\caption{Results (PSNR and SSIM) of the AWGN removal evaluation on the BSD68 dataset. The top two results are emphasized in red and blue, respectively.}
\label{tab:BSD68}
\begin{tabular}{ccccc}
\hline
Metrics & Methods & $\sigma$=15 & $\sigma$=25 & $\sigma$=50\\
\cline{1-5}
\multirow{14}*{PSNR} & BM3D \cite{Dabov2007} & 31.07 & 28.57 & 25.62\\
\cline{2-5}
    & DnCNN-S \cite{Zhang2017} & 31.72 & 29.23 & 26.23\\
\cline{2-5}
    & TNRD \cite{Chen2017} & 31.42 & 28.92 & 25.97\\
\cline{2-5}
    & BUIFD \cite{Helou2020} &  31.35 & 28.75 & 25.11\\
\cline{2-5}
    & IRCNN \cite{ZhangZGZ2017} & 31.63 & 29.15 & 26.19 \\
\cline{2-5}
    & FFDNet \cite{Zhang2018} & 31.63	& 29.19 & 26.29\\
\cline{2-5}
    & DSNetB \cite{Peng2019} & 31.69 & 29.22 & 26.29 \\
\cline{2-5}
    & BRDNet \cite{Tian2020} & \textcolor{blue}{31.79} & 29.29 & 26.36 \\
\cline{2-5}
    & ADNet \cite{TianX2020} & 31.74 & 29.25 & 26.29 \\
\cline{2-5}
    & RIDNet \cite{Anwar2019} & \textcolor{red}{31.81} & \textcolor{blue}{29.34} & \textcolor{blue}{26.40} \\
\cline{2-5}
    & AINDNet \cite{Kim2020} & 31.69 & 29.26 & 26.32 \\
\cline{2-5}
    & DudeNet \cite{Tian2021} & 31.78 & 29.29 & 26.31 \\
\cline{2-5}
    & DRANet & \textcolor{blue}{31.79} & \textcolor{red}{29.36} & \textcolor{red}{26.47}\\
\hline
\multirow{11}*{SSIM} & BM3D \cite{Dabov2007} & 0.872 & 0.802 & 0.687\\
\cline{2-5}
    & TNRD \cite{Chen2017} & 0.883 & 0.816 & 0.703\\
\cline{2-5}
    & DnCNN-S \cite{Zhang2017} & 0.891 & 0.828 & 0.719 \\
\cline{2-5}
    & BUIFD \cite{Helou2020} &  0.886 & 0.819 & 0.682 \\
\cline{2-5}
    & IRCNN  \cite{ZhangZGZ2017} & 0.888 & 0.825 & 0.717 \\
\cline{2-5}
    & FFDNet \cite{Zhang2018} & 0.890 & 0.830 & 0.726 \\
\cline{2-5}
    & BRDNet \cite{Tian2020} & \textcolor{red}{0.893} & \textcolor{blue}{0.831} & \textcolor{blue}{0.727} \\
\cline{2-5}
    & ADNet  \cite{TianX2020} & \textcolor{blue}{0.892} & 0.829 &  0.722 \\
\cline{2-5}
    & RIDNet \cite{Anwar2019} &  \textcolor{red}{0.893} & \textcolor{red}{0.833} & \textcolor{blue}{0.727} \\
\cline{2-5}
    & DRANet & \textcolor{blue}{0.892} & \textcolor{red}{0.833} & \textcolor{red}{0.732} \\
\hline
\end{tabular}
\end{table}

\begin{figure*}[htbp]
	\centering
    \captionsetup[subfigure]{labelformat=empty}
	\begin{subfigure}{0.18\linewidth}
		\centering
		\includegraphics[width=0.99\linewidth]{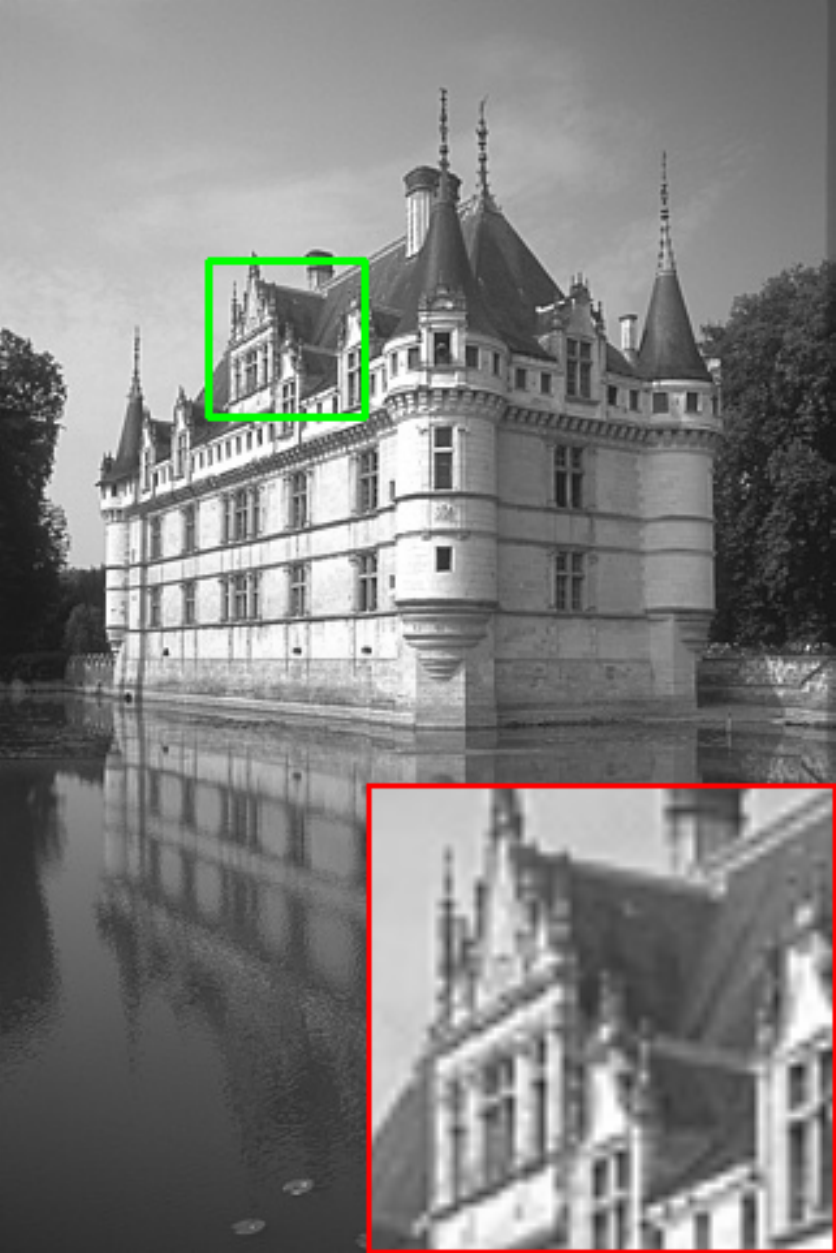}
		\caption{(\romannumeral1)}
	\end{subfigure}
    \centering
	\begin{subfigure}{0.18\linewidth}
		\centering
		\includegraphics[width=0.99\linewidth]{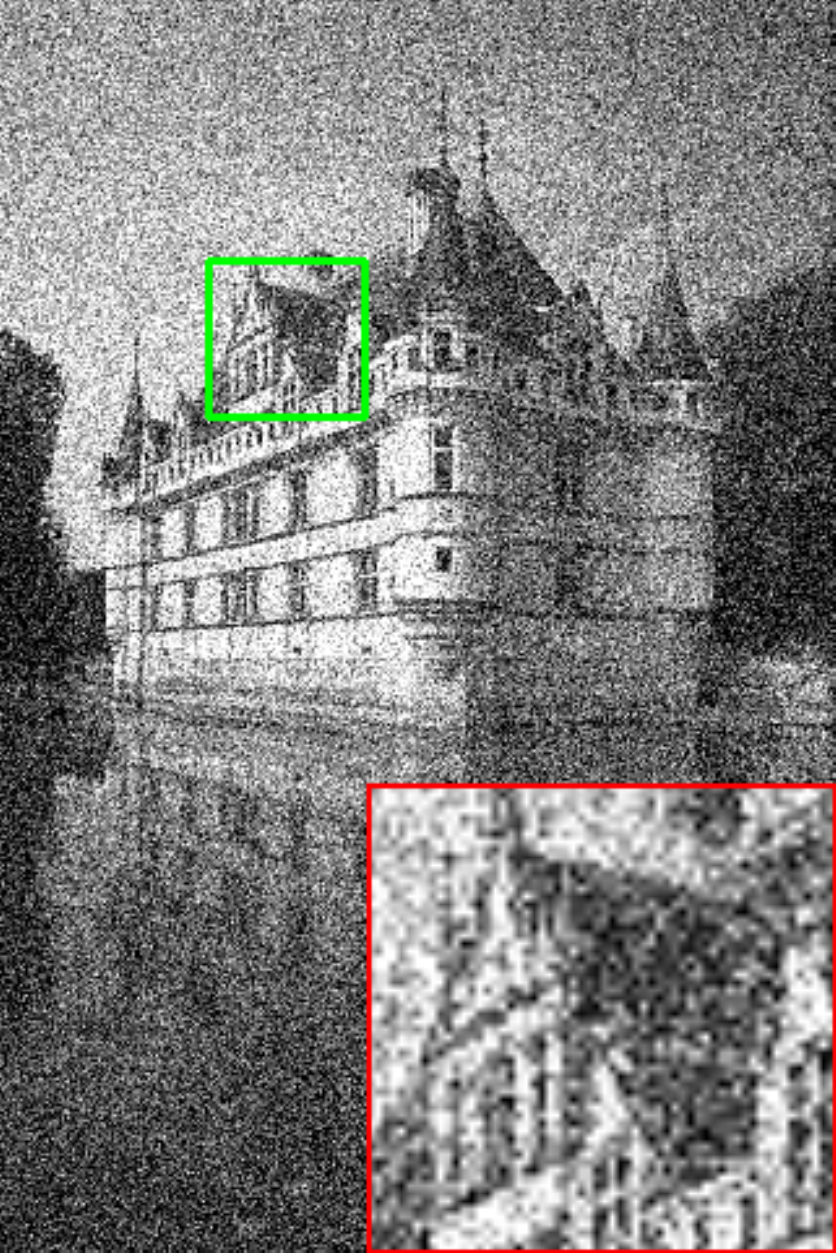}
		\caption{(\romannumeral2)}
	\end{subfigure}
    \centering
	\begin{subfigure}{0.18\linewidth}
		\centering
		\includegraphics[width=0.99\linewidth]{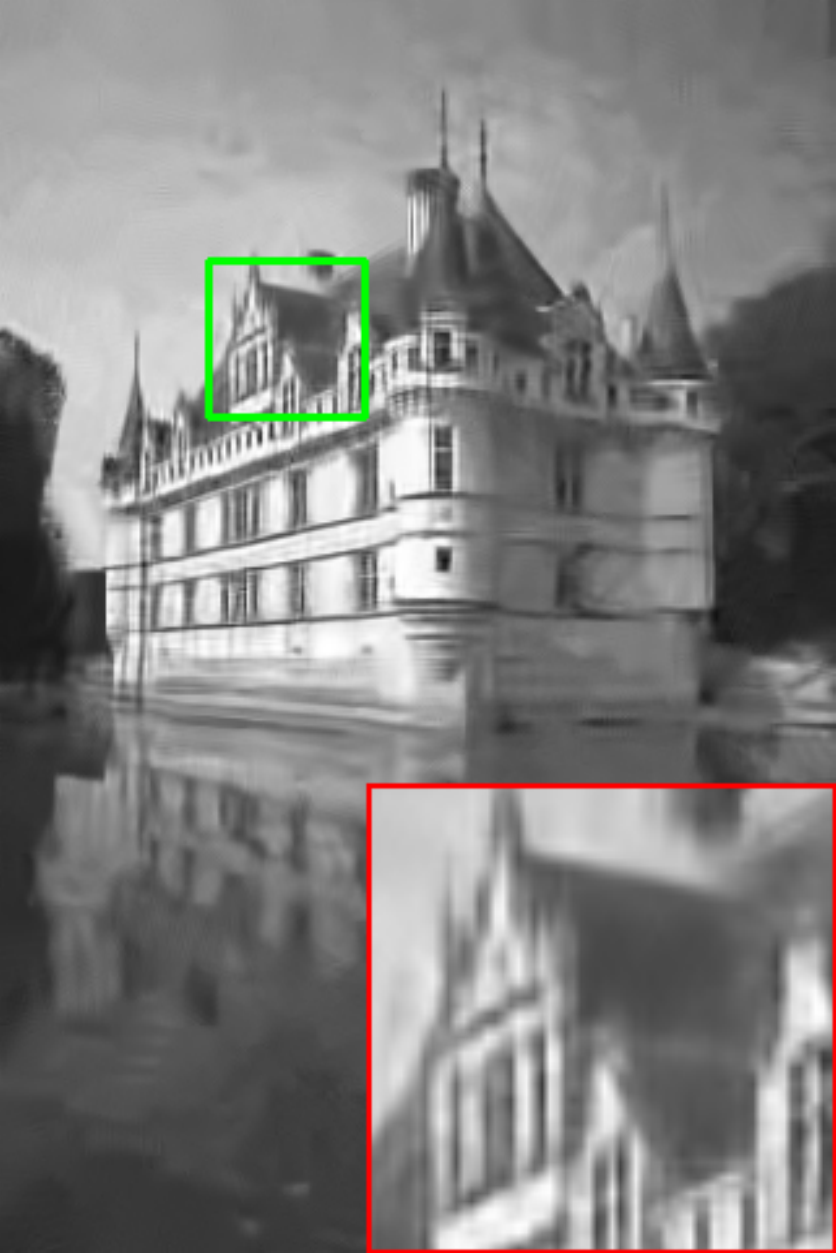}
		\caption{(\romannumeral3)}
	\end{subfigure}
    \centering
	\begin{subfigure}{0.18\linewidth}
		\centering
		\includegraphics[width=0.99\linewidth]{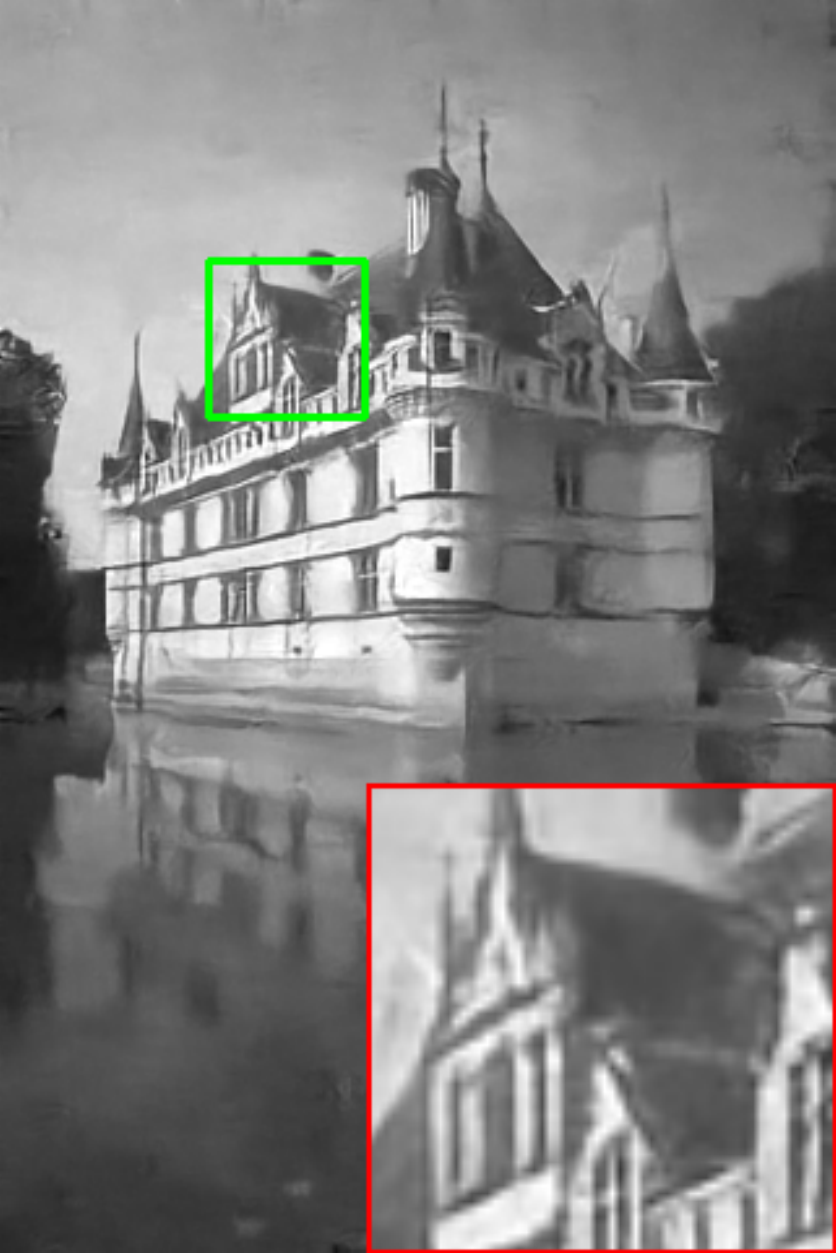}
		\caption{(\romannumeral4)}
	\end{subfigure}
    \centering
	\begin{subfigure}{0.18\linewidth}
		\centering
		\includegraphics[width=0.99\linewidth]{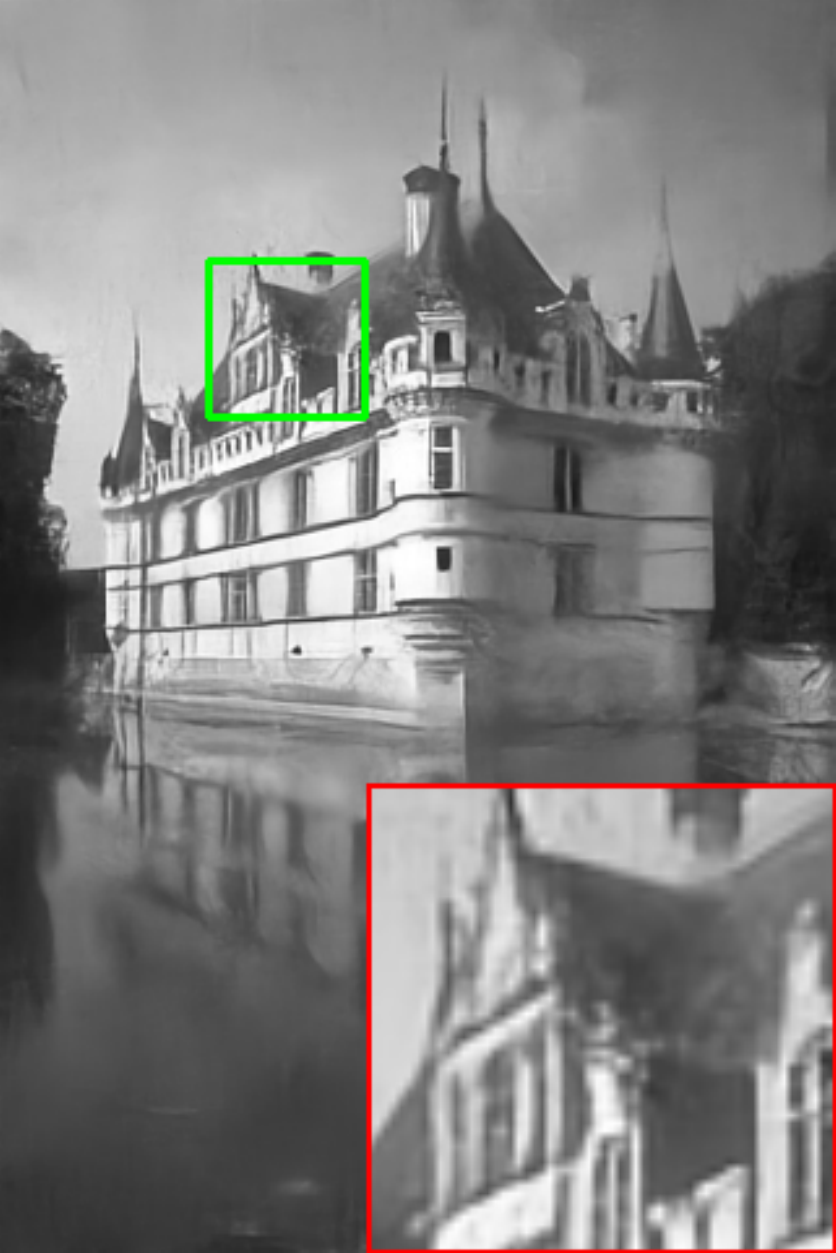}
		\caption{(\romannumeral5)}
	\end{subfigure}
    \centering
	\begin{subfigure}{0.18\linewidth}
		\centering
		\includegraphics[width=0.99\linewidth]{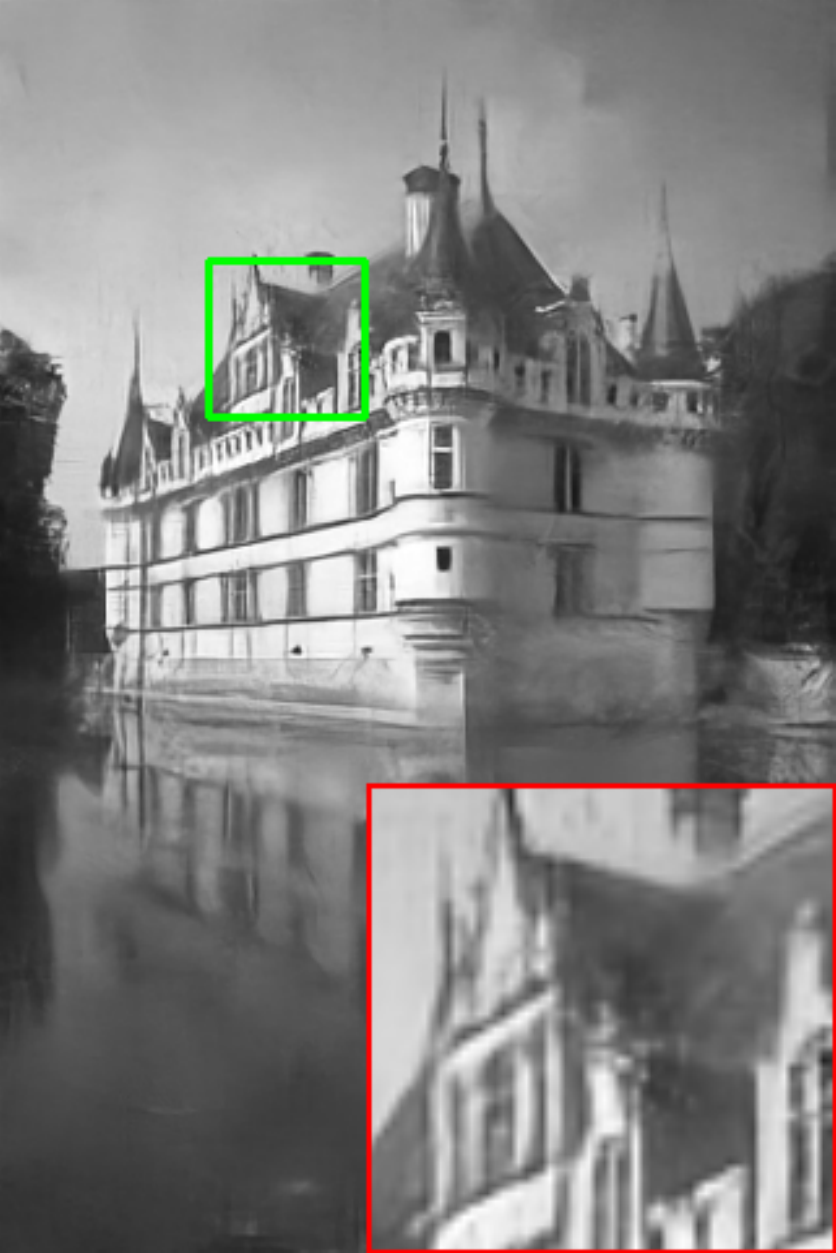}
		\caption{(\romannumeral6)}
	\end{subfigure}
    \centering
	\begin{subfigure}{0.18\linewidth}
		\centering
		\includegraphics[width=0.99\linewidth]{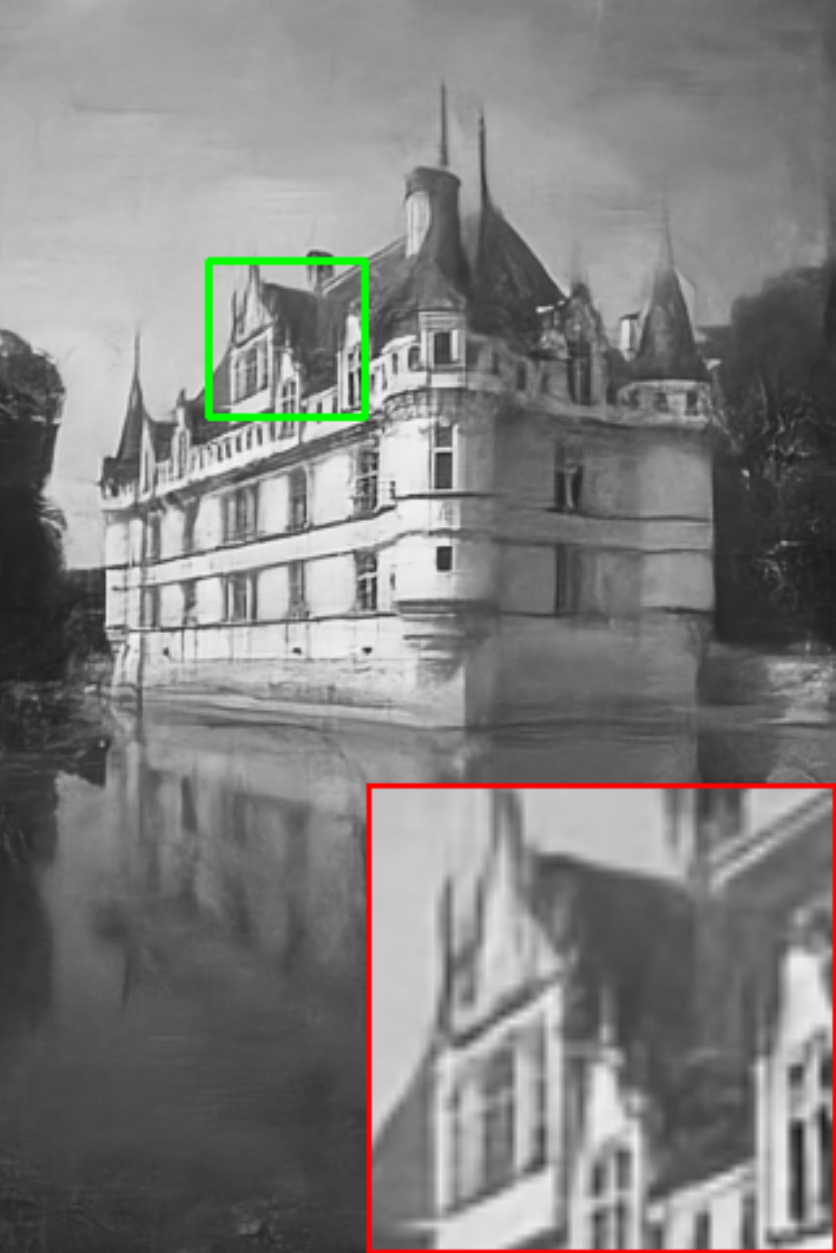}
		\caption{(\romannumeral7)}
	\end{subfigure}
    \centering
	\begin{subfigure}{0.18\linewidth}
		\centering
		\includegraphics[width=0.99\linewidth]{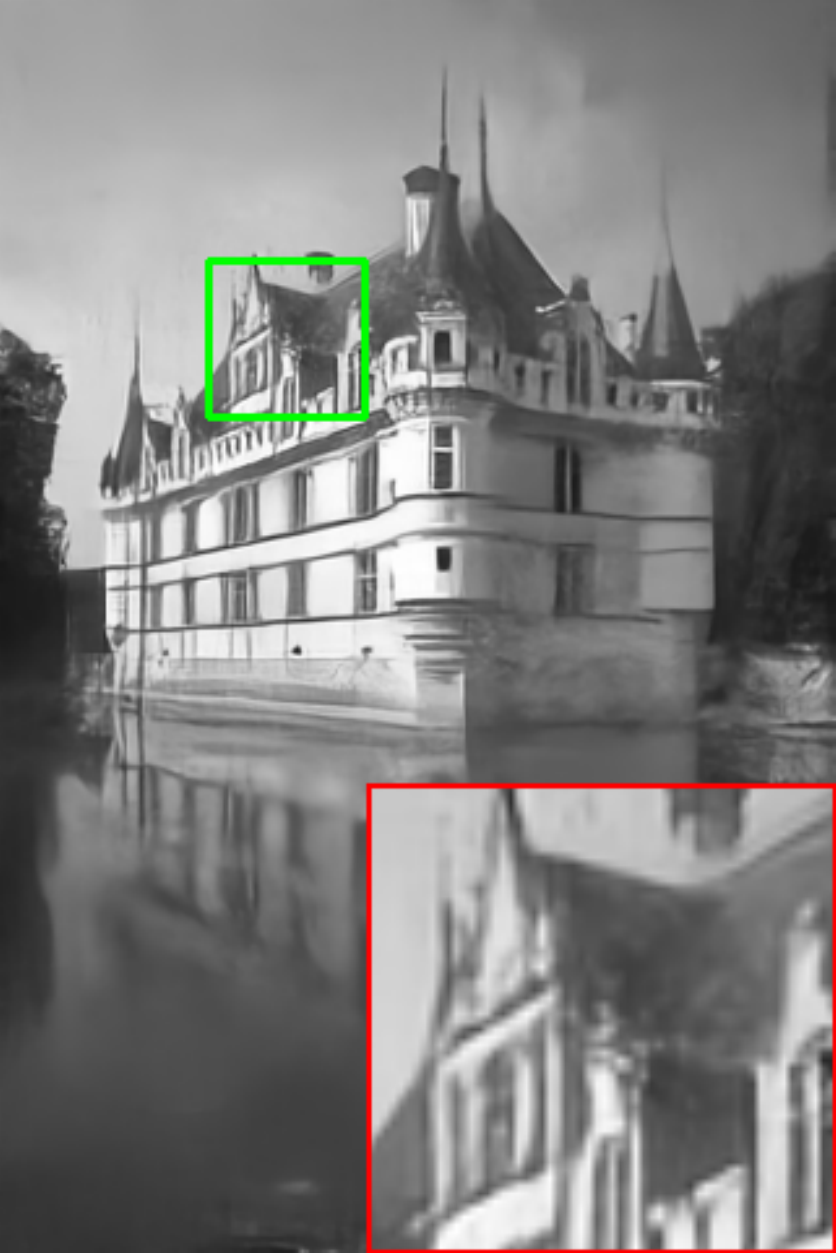}
		\caption{(\romannumeral8)}
	\end{subfigure}
    \centering
	\begin{subfigure}{0.18\linewidth}
		\centering
		\includegraphics[width=0.99\linewidth]{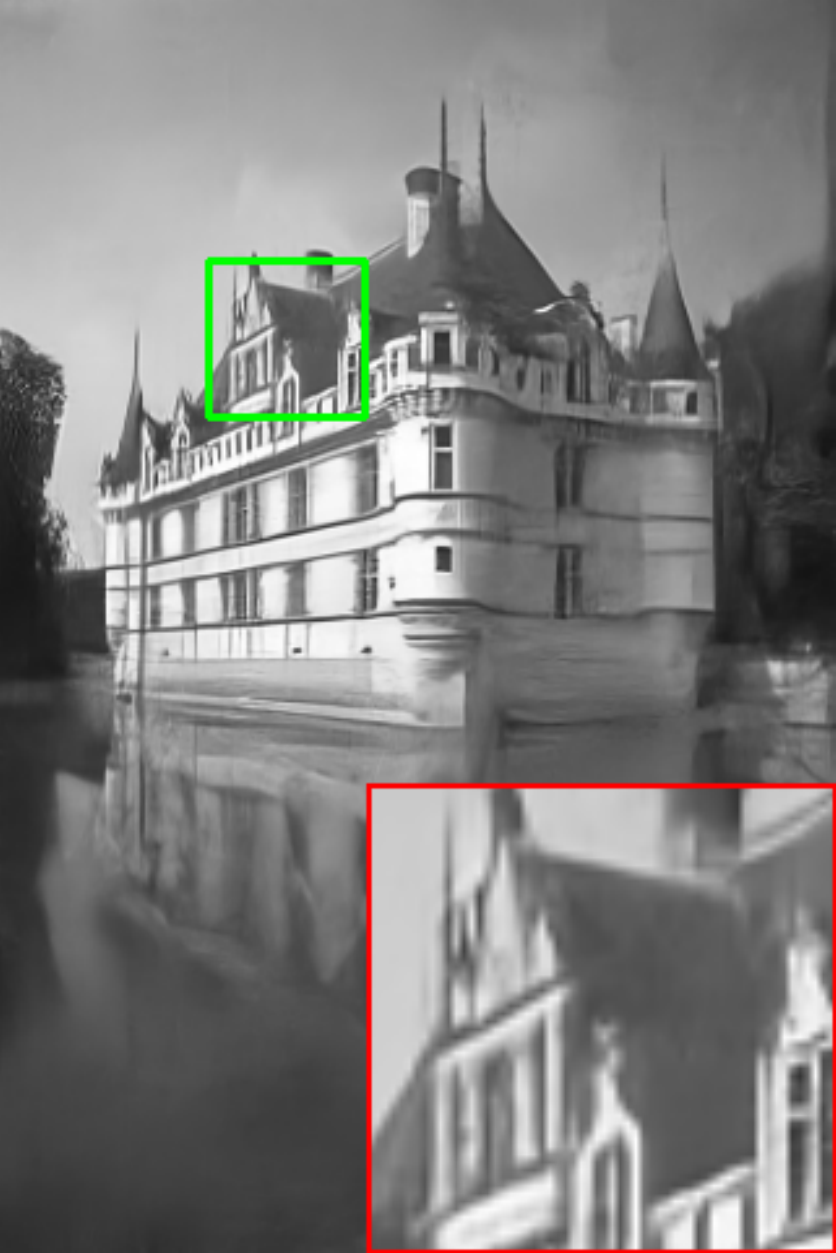}
		\caption{(\romannumeral9)}
	\end{subfigure}
    \centering
	\begin{subfigure}{0.18\linewidth}
		\centering
		\includegraphics[width=0.99\linewidth]{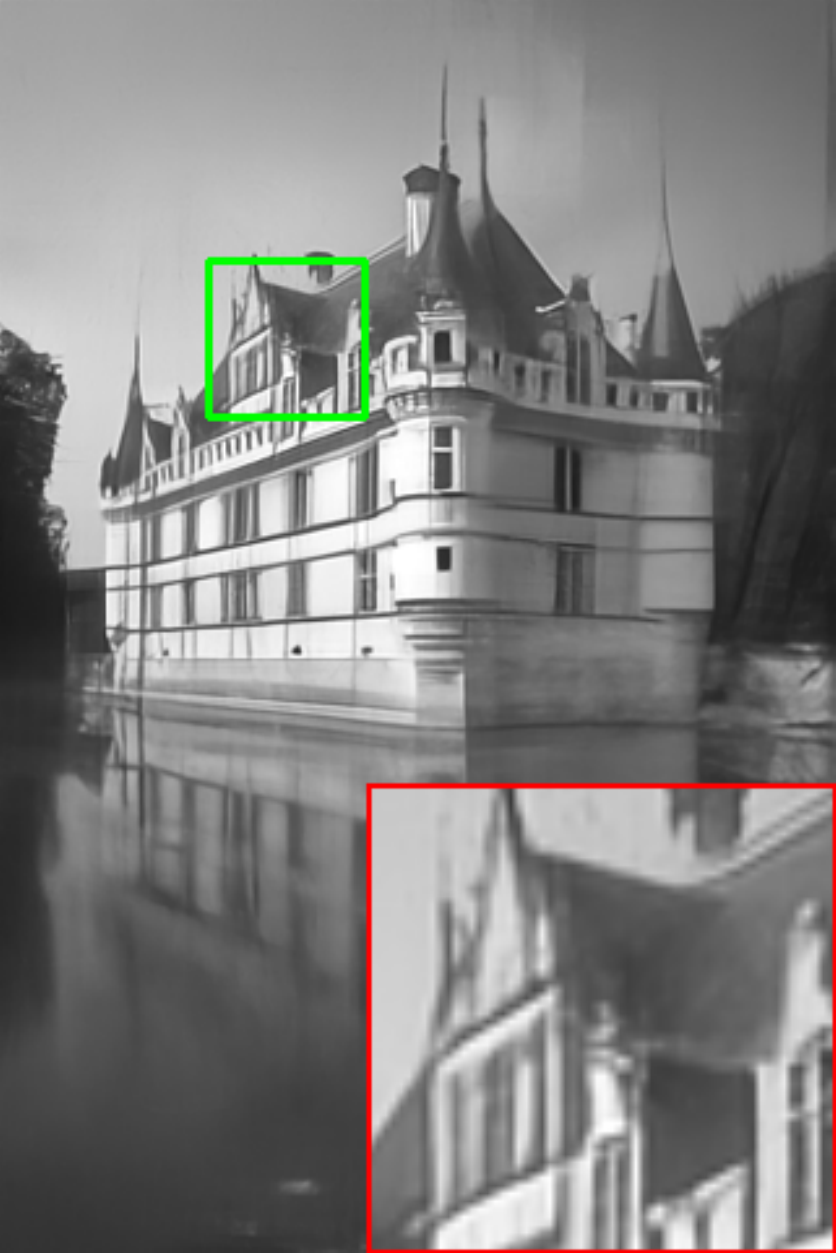}
		\caption{(\romannumeral10)}
	\end{subfigure}
\caption{Visual results of the compared models on the synthetic noisy grayscale image at noise level 50. (\romannumeral1) Ground-truth image, (\romannumeral2) Noisy image / 14.15dB, (\romannumeral3) BM3D / 26.21dB, (\romannumeral4) TNRD / 26.59dB, (\romannumeral5) DnCNN-S / 26.90dB, (\romannumeral6) IRCNN / 26.85dB, (\romannumeral7) BUIFD / 26.32dB, (\romannumeral8) FFDNet / 26.92dB, (\romannumeral9) ADNet / 27.06dB, (\romannumeral10) DRANet / 27.24dB.}
\label{fig:Castle}
\end{figure*}

The visual effects of different models on grayscale image denoising are displayed in Fig. \ref{fig:Castle}, where the ``test003'' image from the BSD68 dataset was selected, and the noise level is set to 50. One can see that the BM3D, TNRD, DnCNN-S, and IRCNN remove noise to some extent, however they obtained blurred results. The BUIFD and FFDNet lost many textures and details of the image. The ADNet produces the over-smoothed edges. In contrast, the DRANet obtains a visual appealing result, it also obtains the best PSNR value among the compared models. Namely, the DRANet can achieve better results both subjectively and objectively.

The proposed model was also tested for synthetic noise removal of color images. Three color image datasets: the CBSD68 \cite{Roth2005}, Kodak24 \cite{Kodak24}, and McMaster \cite{Zhang2011} datasets, were employed for the color image denoising test. The average PSNR and SSIM values of the compared models are listed in Table \ref{tab:color_PSNR} and Table \ref{tab:color_SSIM}, respectively. Compared with other methods, the DRANet obtained the leading denoising performances at all noise levels on the three datasets, both on the average PSNR and SSIM.

\begin{table*}[htbp]
\centering
\caption{The evaluation results (PSNR) of the AWGN removal on color images. The top two results are highlighted in red and blue, respectively.}
\label{tab:color_PSNR}
\begin{tabular}{ccccc}
\cline{1-5}
Datasets &  Methods & $\sigma$=15 & $\sigma$=25 & $\sigma$=50 \\
\cline{1-5}
\multirow{13}*{CBSD68} & CBM3D \cite{Dabov2007} & 33.52 & 30.71 & 27.38\\
\cline{2-5}
    & CDnCNN-S \cite{Zhang2017} & 33.89 & 31.23 & 27.92 \\
\cline{2-5}
    & BUIFD \cite{Helou2020} & 33.65 & 30.76 & 26.61 \\
\cline{2-5}
    & IRCNN \cite{ZhangZGZ2017} & 33.86 & 31.16 & 27.86\\
\cline{2-5}
    & FFDNet \cite{Zhang2018} & 33.87 & 31.21 & 27.96\\
\cline{2-5}
    & DSNetB \cite{Peng2019} & 33.91 & 31.28 & 28.05\\
\cline{2-5}
    & RIDNet \cite{Anwar2019} & 34.01 & 31.37 & 28.14\\
\cline{2-5}
    & VDN \cite{Yue2019} & 33.90 & 31.35 & 28.19\\
\cline{2-5}
    & BRDNet \cite{Tian2020} & \textcolor{blue}{34.10} & \textcolor{blue}{31.43} & \textcolor{blue}{28.16} \\
\cline{2-5}
    & ADNet \cite{TianX2020} & 33.99 & 31.31 & 28.04\\
\cline{2-5}
    & DudeNet \cite{Tian2021} & 34.01 & 31.34 & 28.09\\
\cline{2-5}
    & AirNet \cite{Li2022} & 33.92 & 31.26 & 28.01\\
\cline{2-5}
    & DRANet & \textcolor{red}{34.18} & \textcolor{red}{31.56} & \textcolor{red}{28.37}\\
\cline{1-5}
\multirow{12}*{Kodak24} & CBM3D \cite{Dabov2007} & 34.28 & 31.68 & 28.46\\
\cline{2-5}
    & CDnCNN-S \cite{Zhang2017} & 34.48 & 32.03 & 28.85 \\
\cline{2-5}
    & BUIFD \cite{Helou2020} & 34.41 & 31.77 & 27.74\\
\cline{2-5}
    & IRCNN \cite{ZhangZGZ2017} & 34.56 & 32.03 & 28.81\\
\cline{2-5}
    & FFDNet \cite{Zhang2018} & 34.63 & 32.13 & 28.98 \\
\cline{2-5}
    & DSNetB \cite{Peng2019} & 34.63 & 32.16 & 29.05 \\
\cline{2-5}
    & BRDNet \cite{Tian2020} & \textcolor{blue}{34.88} & \textcolor{blue}{32.41} & \textcolor{blue}{29.22} \\
\cline{2-5}
    & ADNet \cite{TianX2020} & 34.76 & 32.26 & 29.10 \\
\cline{2-5}
    & DudeNet \cite{Tian2021} & 34.81 & 32.26 & 29.10 \\
\cline{2-5}
    & AirNet \cite{Li2022} & 34.68 & 32.21 & 29.06\\
\cline{2-5}
    & DRANet & \textcolor{red}{35.02} & \textcolor{red}{32.59} & \textcolor{red}{29.50} \\
\cline{1-5}
\multirow{11}*{McMaster} & CBM3D \cite{Dabov2007} & 34.06 & 31.66 & 28.51 \\
\cline{2-5}
    & CDnCNN-S \cite{Zhang2017} & 33.44 & 31.51 & 28.61\\
\cline{2-5}
    & BUIFD \cite{Helou2020} & 33.84 & 31.06 & 26.20\\
\cline{2-5}
    & IRCNN \cite{ZhangZGZ2017} & 34.58 & 32.18 & 28.91\\
\cline{2-5}
    & FFDNet \cite{Zhang2018} & 34.66 & 32.35 & 29.18\\
\cline{2-5}
    & DSNetB \cite{Peng2019} & 34.67 & 32.40 & 29.28\\
\cline{2-5}
    & BRDNet \cite{Tian2020} & \textcolor{blue}{35.08} & \textcolor{blue}{32.75} & \textcolor{blue}{29.52} \\
\cline{2-5}
    & ADNet \cite{TianX2020} & 34.93 & 32.56  & 29.36 \\
\cline{2-5}
    & AirNet \cite{Li2022} & 34.70 & 32.44 & 29.26 \\
\cline{2-5}
    & DRANet & \textcolor{red}{35.09} & \textcolor{red}{32.84} & \textcolor{red}{29.77} \\
\cline{1-5}
\end{tabular}
\end{table*}

\begin{table}[htbp]
\centering
\caption{The evaluation results (SSIM) of the AWGN removal on color images. The top two results are highlighted in red and blue, respectively.}
\label{tab:color_SSIM}
\begin{tabular}{ccccc}
\hline
Datasets & Methods & $\sigma$=15 & $\sigma$=25 & $\sigma$=50\\
\cline{1-5}
\multirow{7}*{CBSD68} & CDnCNN-B \cite{Zhang2017} & 0.929 & 0.883 & 0.790\\
\cline{2-5}
    & BUIFD \cite{Helou2020} & \textcolor{blue}{0.930} & 0.882 & 0.777\\
\cline{2-5}
    & IRCNN  \cite{ZhangZGZ2017} & 0.929 & 0.882 & 0.790\\
\cline{2-5}
    & FFDNet \cite{Zhang2018} & 0.929 & 0.882 & 0.789 \\
\cline{2-5}
    & ADNet \cite{TianX2020}  & \textcolor{red}{0.933} & \textcolor{blue}{0.889} & 0.797\\
\cline{2-5}
    & AirNet \cite{Li2022} & \textcolor{red}{0.933} & 0.888 & \textcolor{blue}{0.798}\\
\cline{2-5}
    & DRANet & \textcolor{red}{0.933} & \textcolor{red}{0.890} & \textcolor{red}{0.805}\\
\cline{1-5}
\multirow{7}*{Kodak24} & CDnCNN-B \cite{Zhang2017} & 0.920 & 0.876 & 0.791\\
\cline{2-5}
    & BUIFD \cite{Helou2020} & 0.923 & 0.879 & 0.786\\
\cline{2-5}
    & IRCNN  \cite{ZhangZGZ2017} & 0.920 & 0.877 & 0.793\\
\cline{2-5}
    & FFDNet \cite{Zhang2018} & 0.922 & 0.878 & 0.794\\
\cline{2-5}
    & ADNet \cite{TianX2020}  & \textcolor{blue}{0.924} & \textcolor{blue}{0.882} & 0.798\\
\cline{2-5}
    & AirNet \cite{Li2022} & \textcolor{blue}{0.924} & \textcolor{blue}{0.882} & \textcolor{blue}{0.799}\\
\cline{2-5}
    & DRANet & \textcolor{red}{0.927} & \textcolor{red}{0.888} & \textcolor{red}{0.813}\\
\cline{1-5}
\multirow{7}*{McMaster} & CDnCNN-B \cite{Zhang2017} & 0.904 & 0.869 & 0.799\\
\cline{2-5}
    & BUIFD \cite{Helou2020} & 0.901 & 0.847 & 0.733\\
\cline{2-5}
    & IRCNN  \cite{ZhangZGZ2017} & 0.920 & 0.882 & 0.807\\
\cline{2-5}
    & FFDNet \cite{Zhang2018} & 0.922 & 0.886 & 0.815 \\
\cline{2-5}
    & ADNet \cite{TianX2020}  & \textcolor{red}{0.927} & \textcolor{blue}{0.894} & \textcolor{blue}{0.825}\\
\cline{2-5}
    & AirNet \cite{Li2022} & \textcolor{blue}{0.925} & 0.891 & 0.822\\
\cline{2-5}
    & DRANet & \textcolor{red}{0.927} & \textcolor{red}{0.896} & \textcolor{red}{0.835}\\
\cline{1-5}
\end{tabular}
\end{table}

\begin{figure*}[htbp]
	\centering
    \captionsetup[subfigure]{labelformat=empty}
	\begin{subfigure}{0.225\linewidth}
		\centering
		\includegraphics[width=0.99\linewidth]{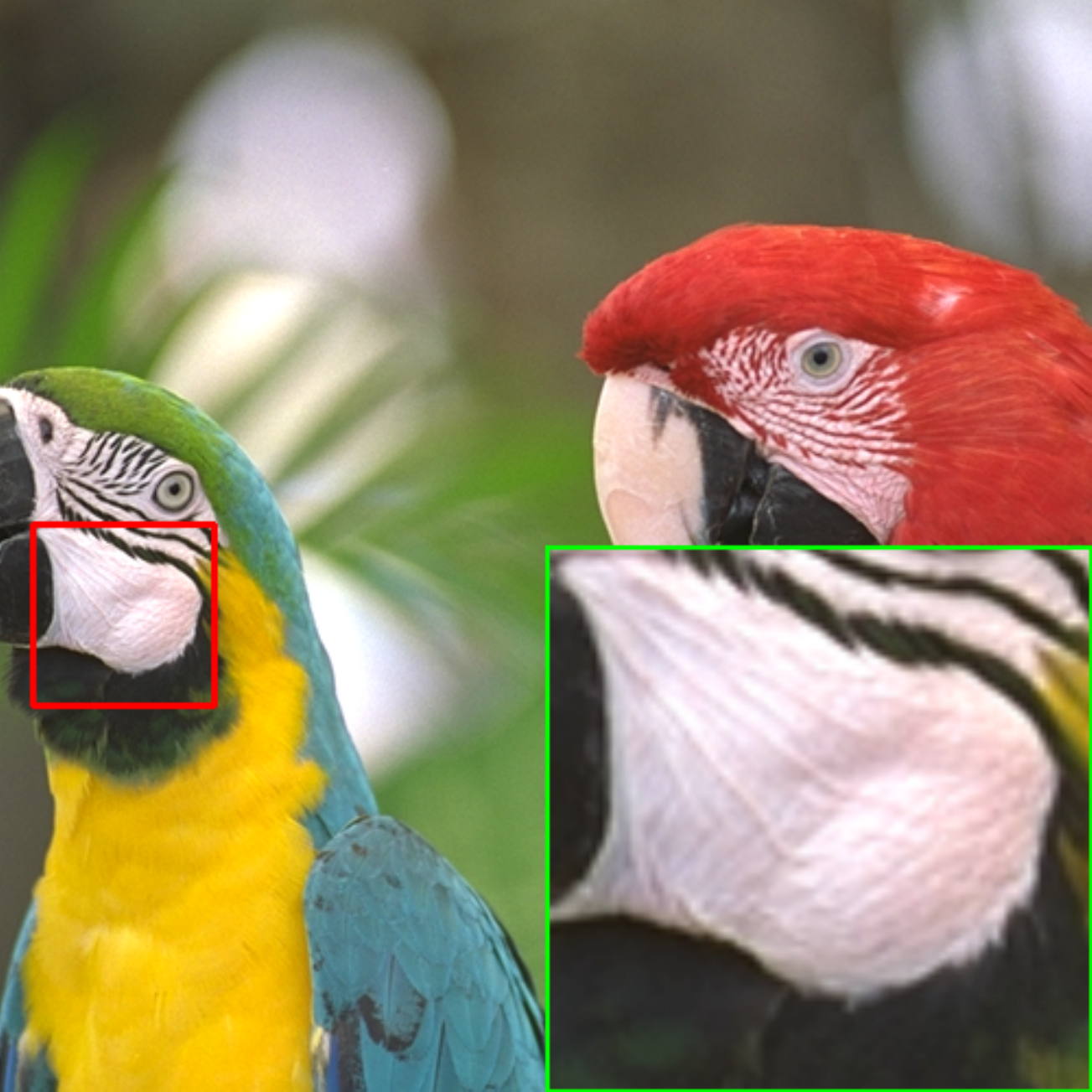}
		\caption{(\romannumeral1)}
	\end{subfigure}
    \centering
	\begin{subfigure}{0.225\linewidth}
		\centering
		\includegraphics[width=0.99\linewidth]{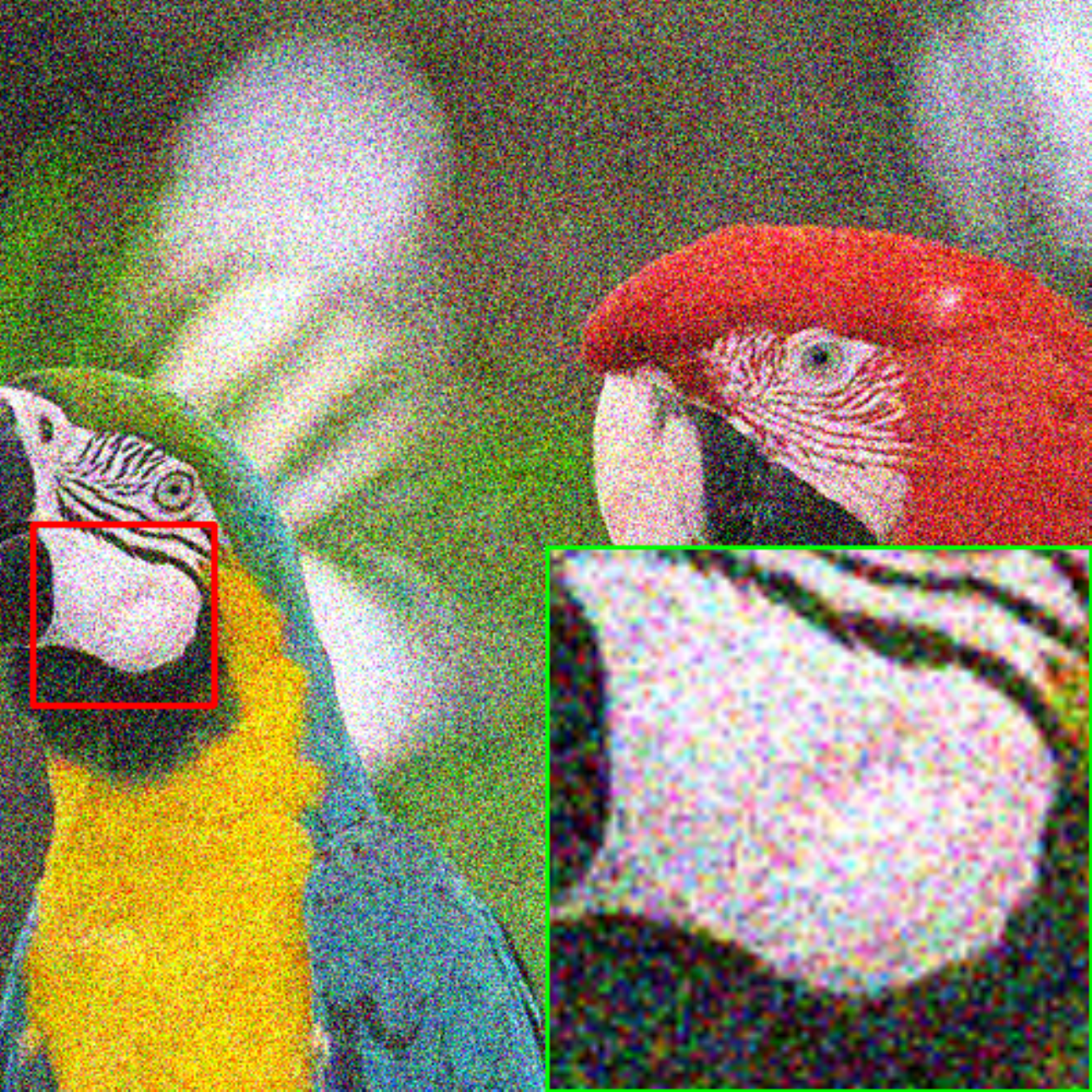}
		\caption{(\romannumeral2)}
	\end{subfigure}
    \centering
	\begin{subfigure}{0.225\linewidth}
		\centering
		\includegraphics[width=0.99\linewidth]{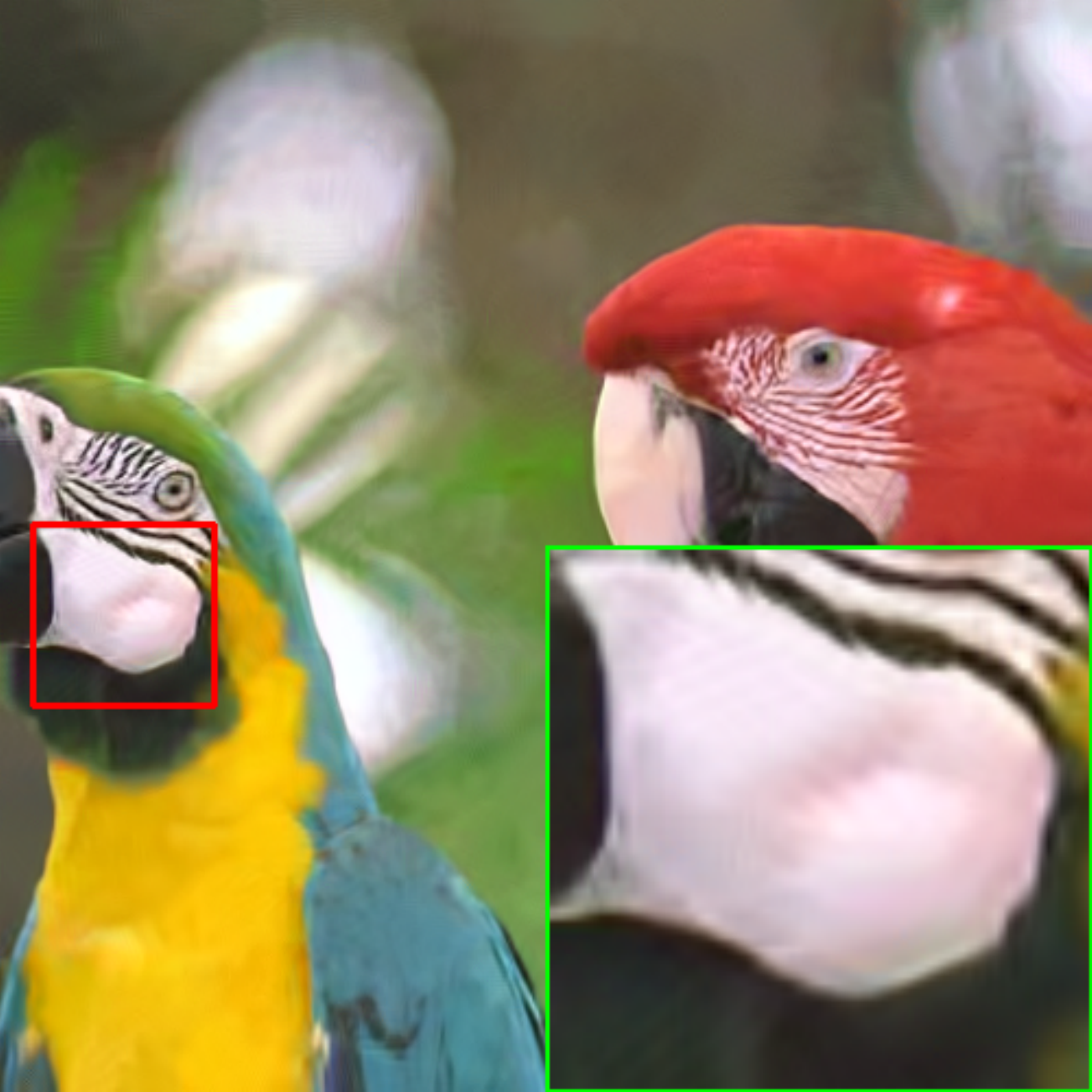}
		\caption{(\romannumeral3)}
	\end{subfigure}
    \centering
	\begin{subfigure}{0.225\linewidth}
		\centering
		\includegraphics[width=0.99\linewidth]{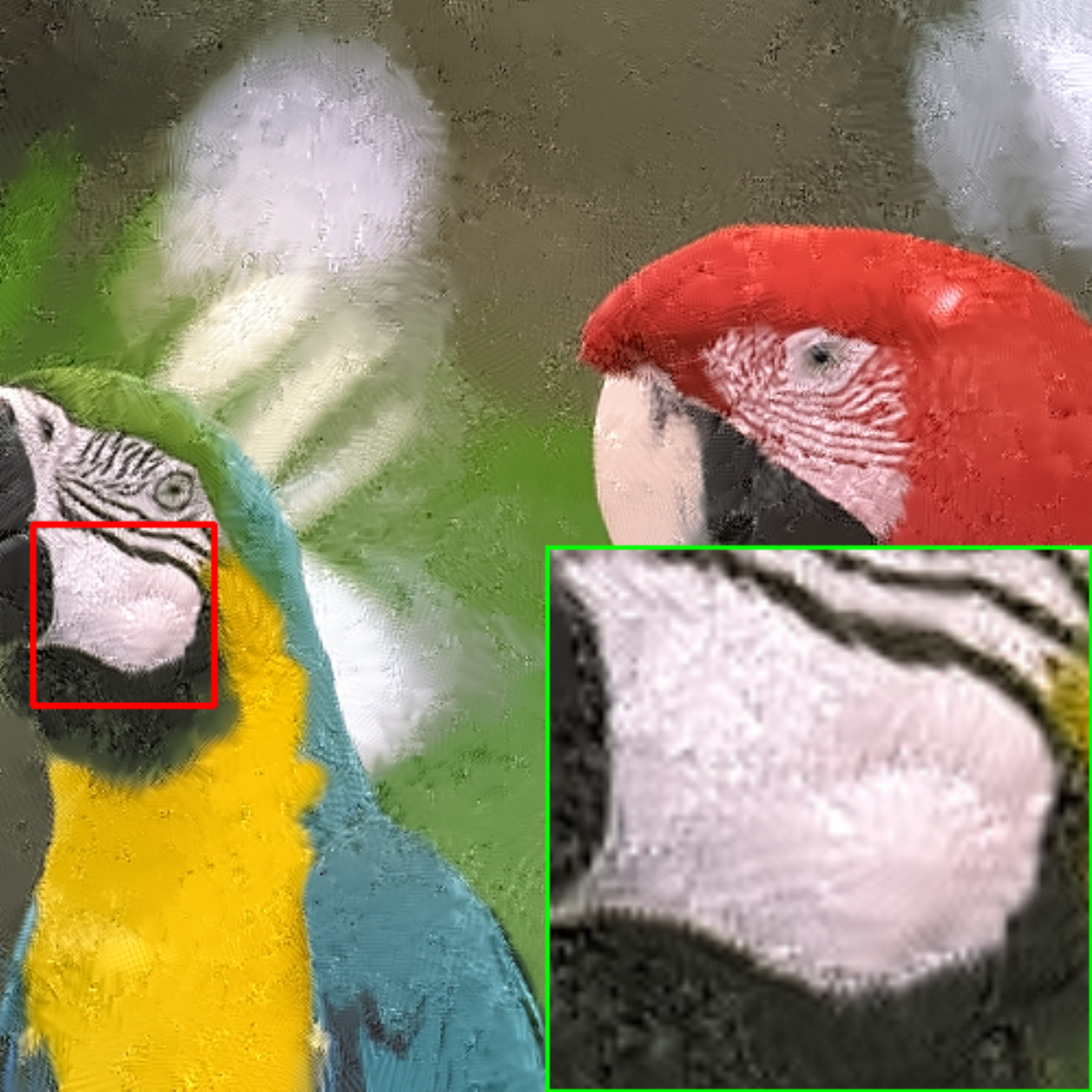}
		\caption{(\romannumeral4)}
	\end{subfigure}
    \centering
	\begin{subfigure}{0.225\linewidth}
		\centering
		\includegraphics[width=0.99\linewidth]{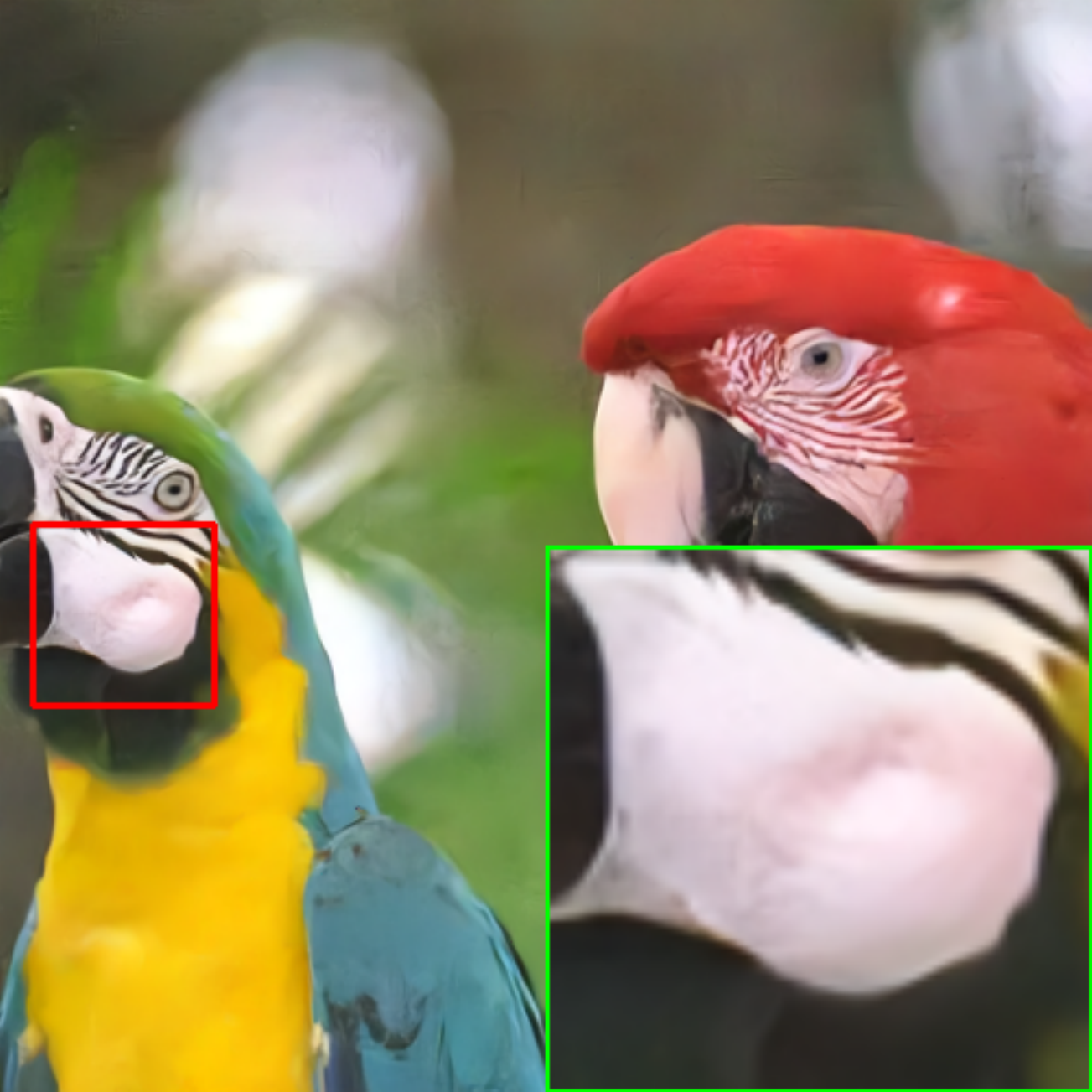}
		\caption{(\romannumeral5)}
	\end{subfigure}
    \centering
    \begin{subfigure}{0.225\linewidth}
		\centering
		\includegraphics[width=0.99\linewidth]{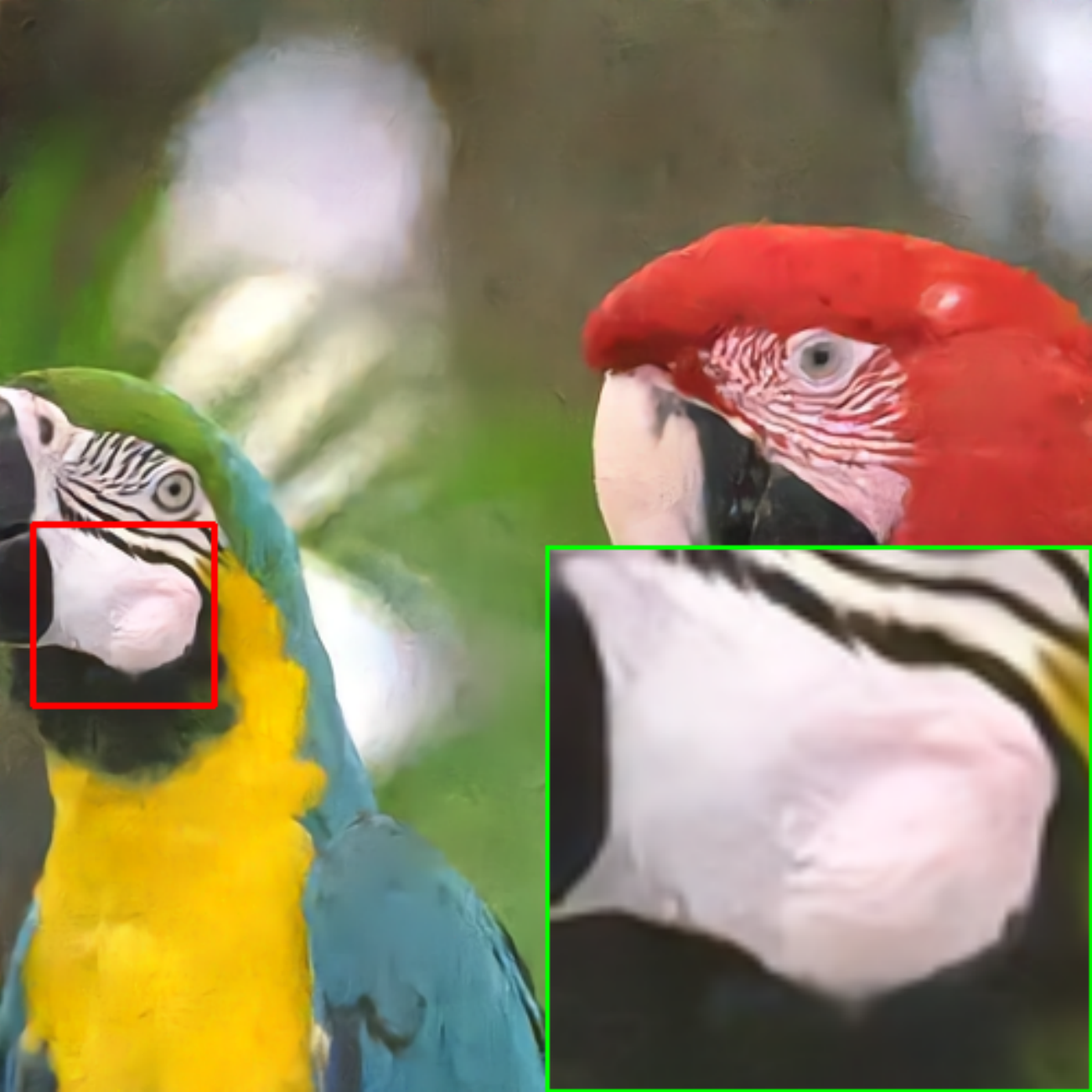}
		\caption{(\romannumeral6)}
	\end{subfigure}
    \centering
	\begin{subfigure}{0.225\linewidth}
		\centering
		\includegraphics[width=0.99\linewidth]{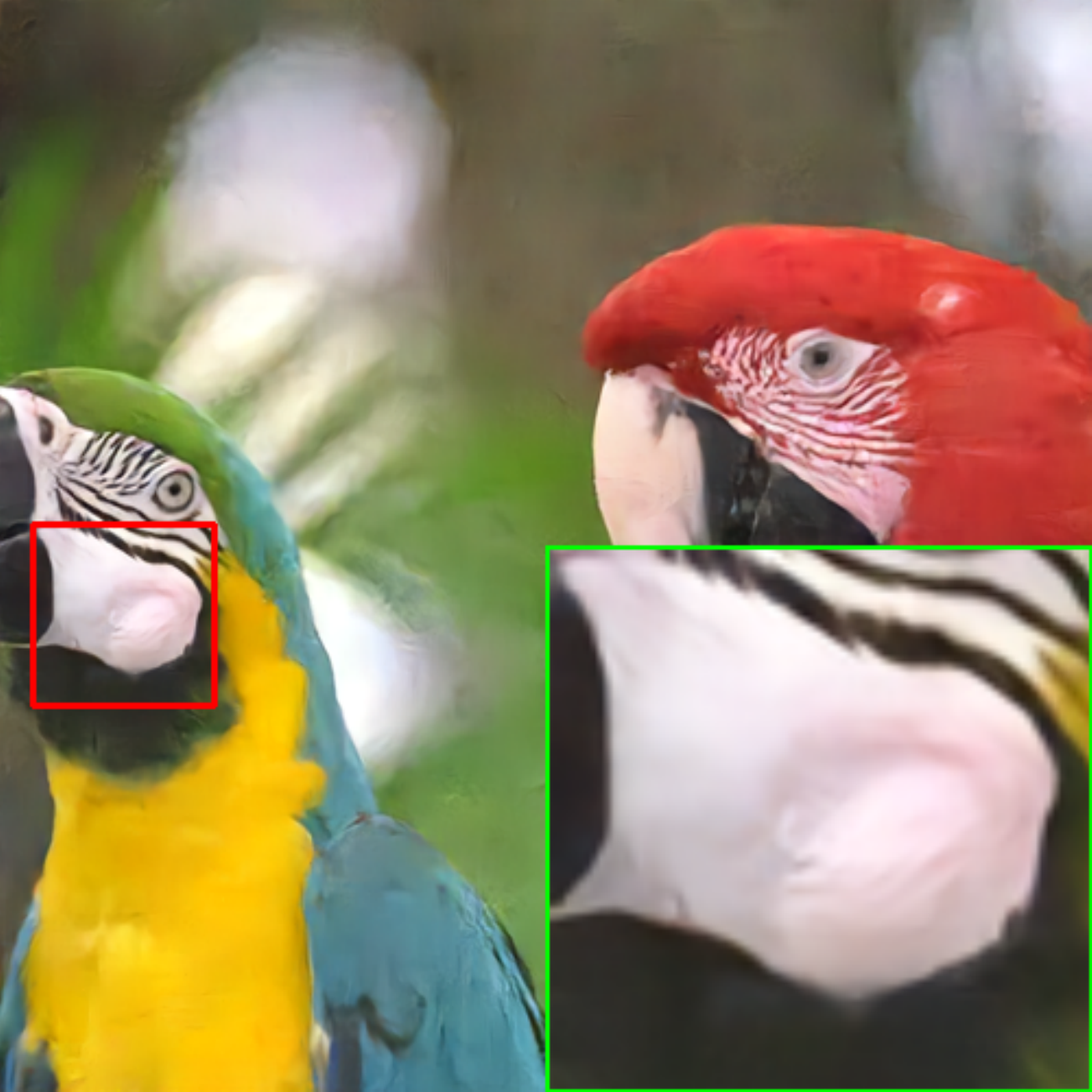}
		\caption{(\romannumeral7)}
	\end{subfigure}
    \centering
	\begin{subfigure}{0.225\linewidth}
		\centering
		\includegraphics[width=0.99\linewidth]{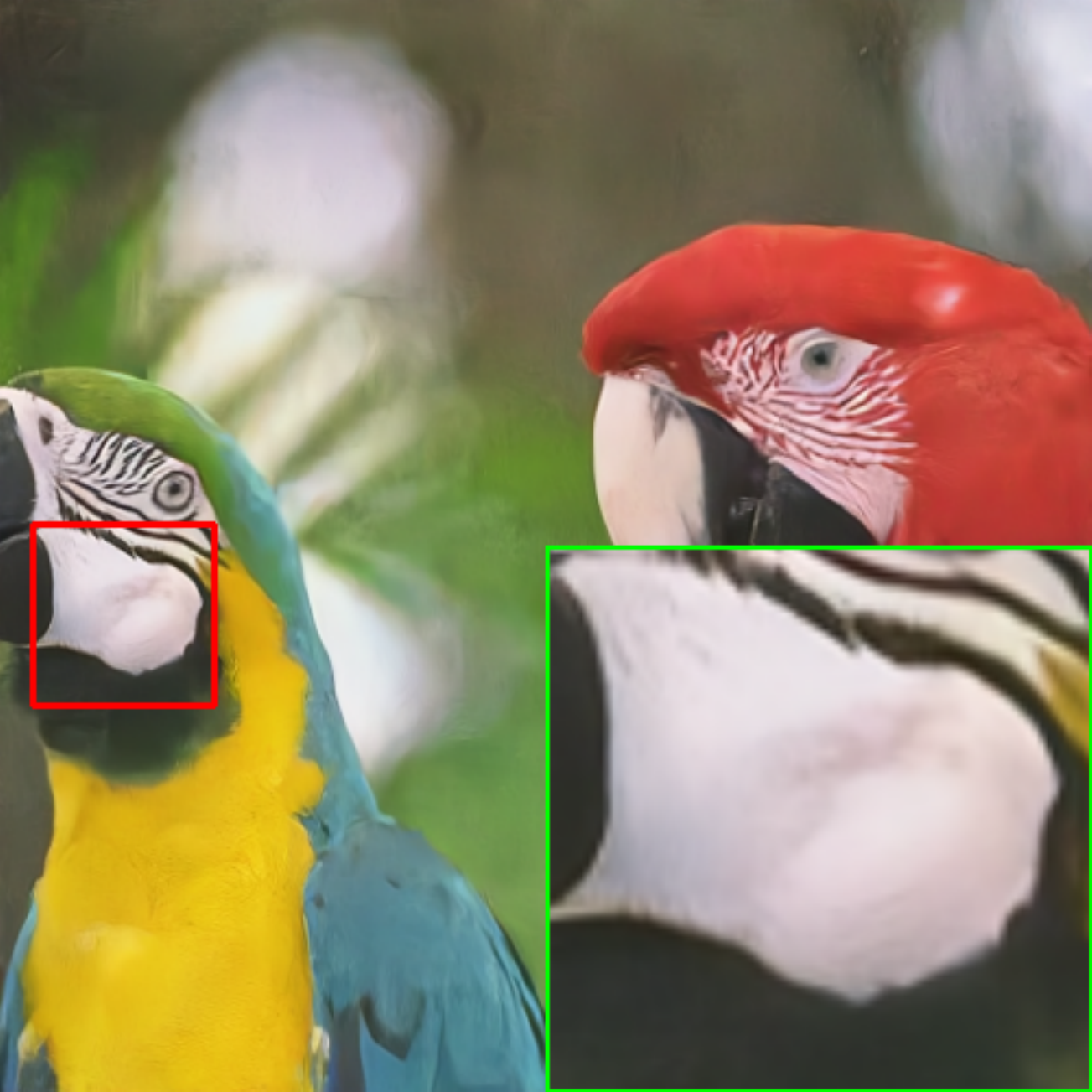}
		\caption{(\romannumeral8)}
	\end{subfigure}
    \centering
	\begin{subfigure}{0.225\linewidth}
		\centering
		\includegraphics[width=0.99\linewidth]{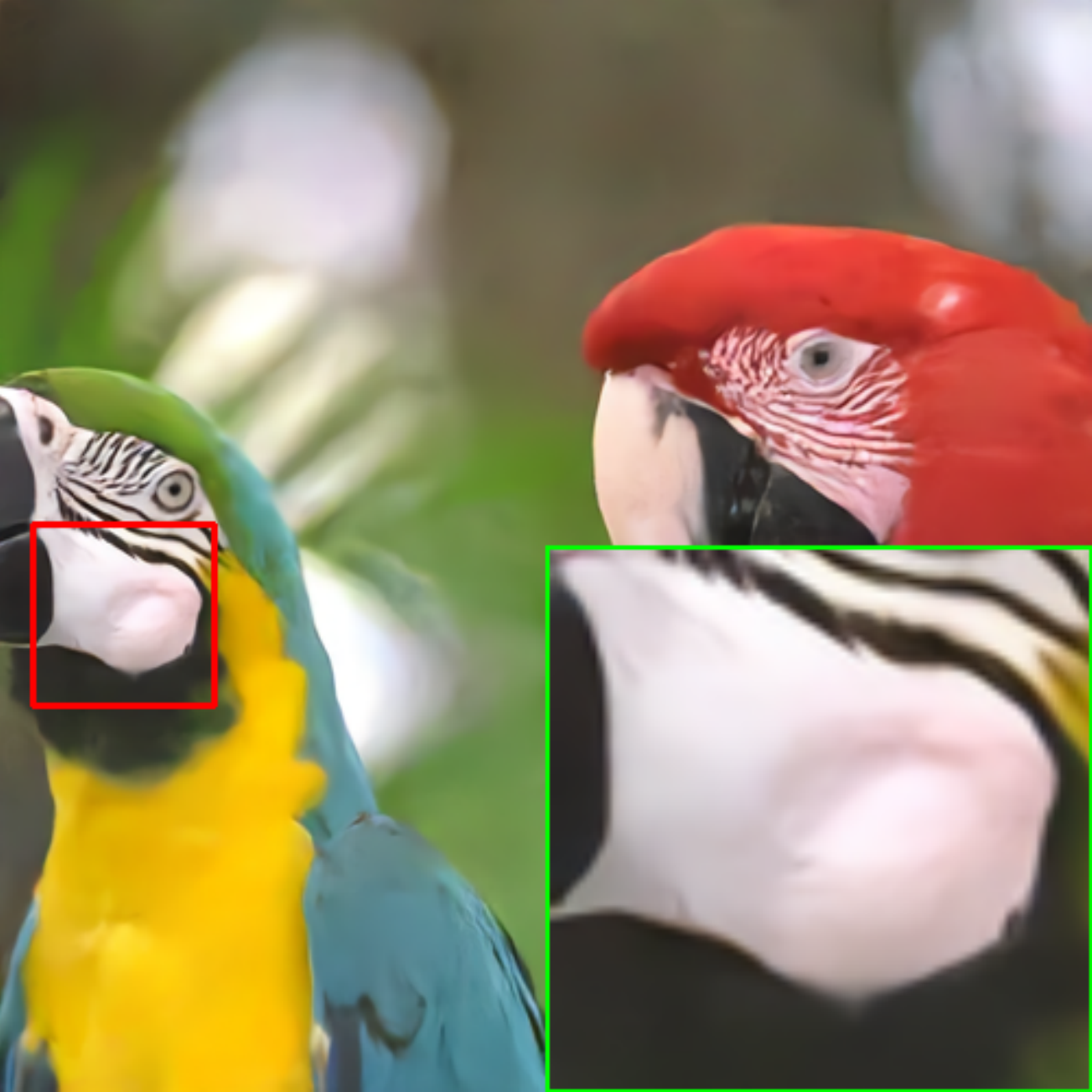}
		\caption{(\romannumeral9)}
	\end{subfigure}
    \centering
	\begin{subfigure}{0.225\linewidth}
		\centering
		\includegraphics[width=0.99\linewidth]{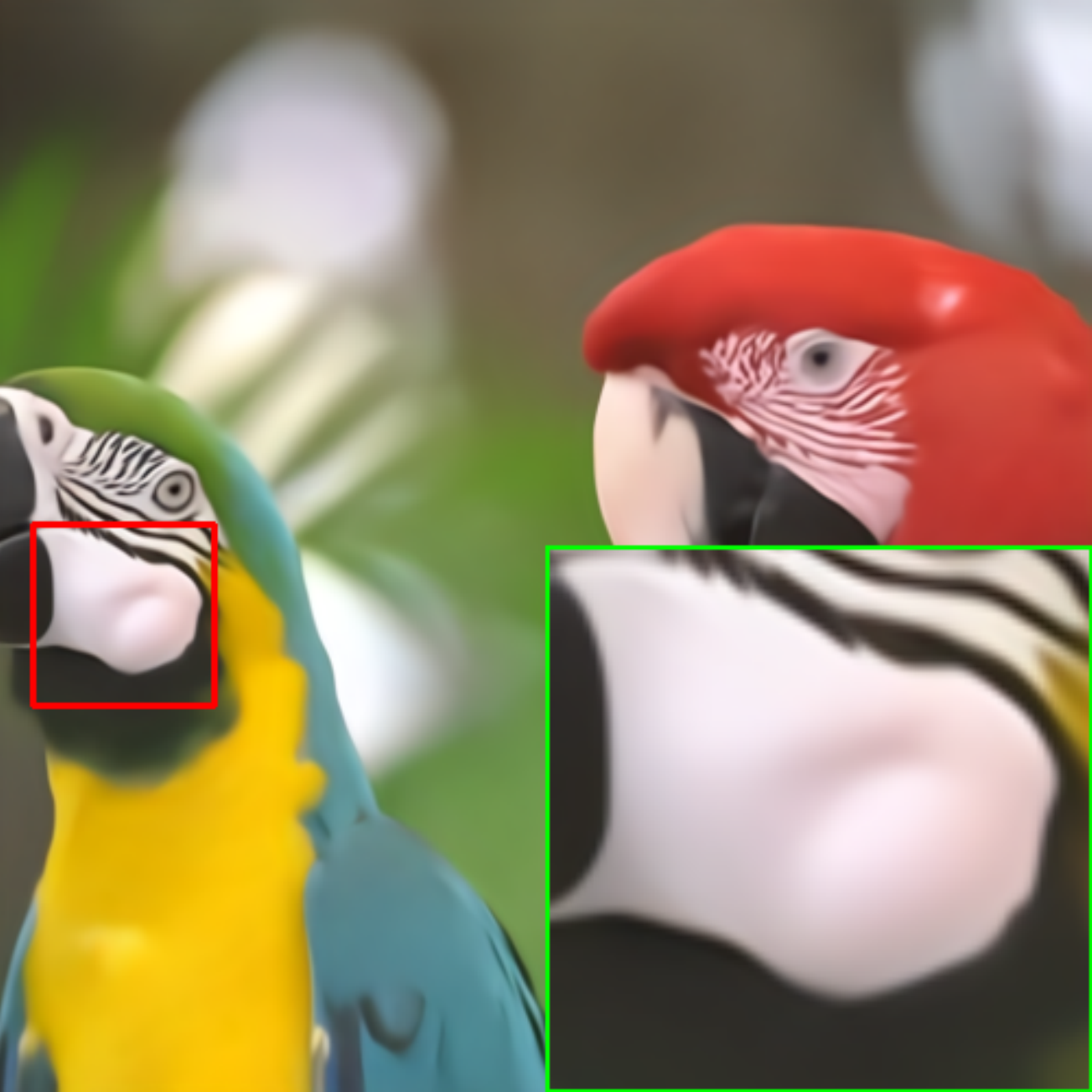}
		\caption{(\romannumeral10)}
	\end{subfigure}
    \centering
	\begin{subfigure}{0.225\linewidth}
		\centering
		\includegraphics[width=0.99\linewidth]{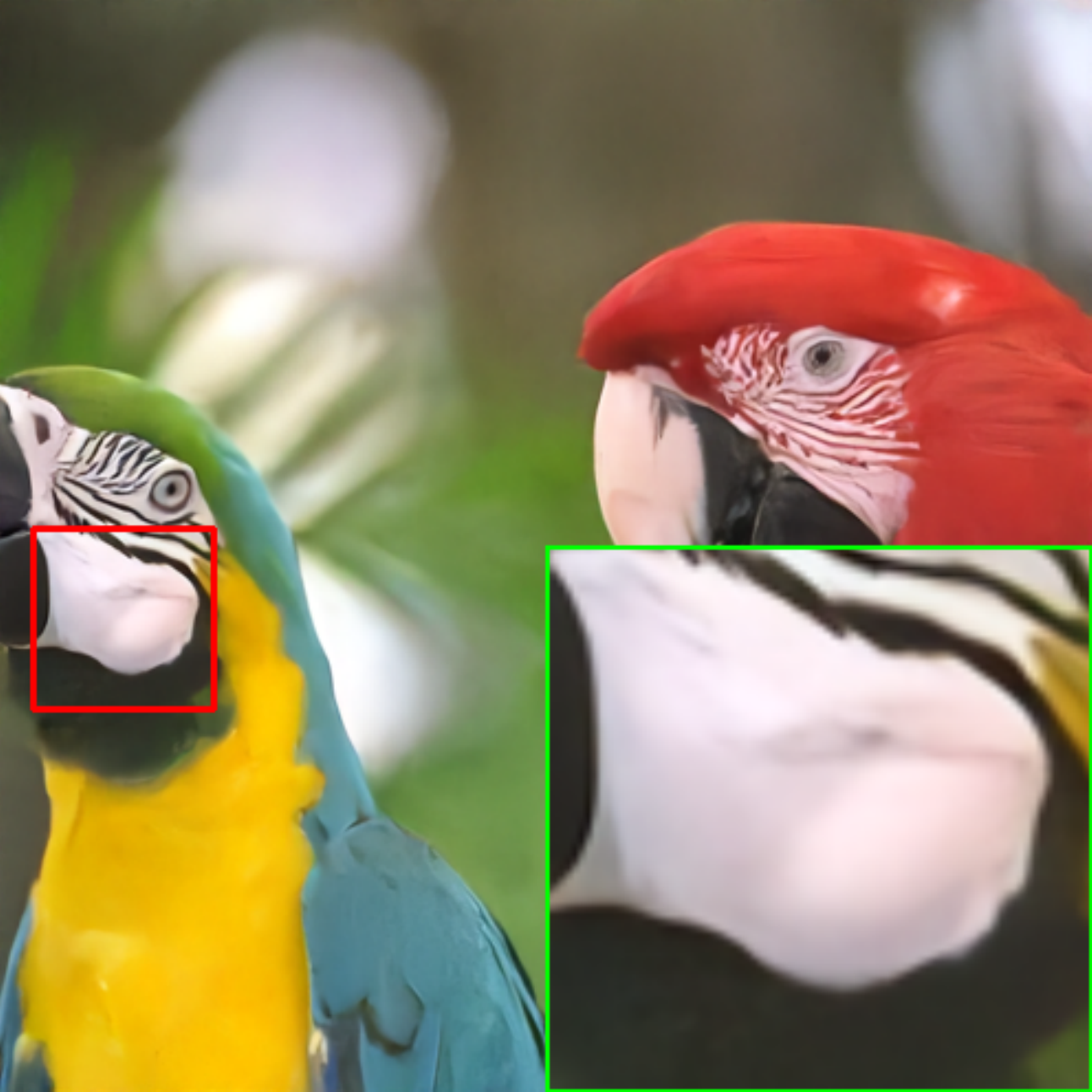}
		\caption{(\romannumeral11)}
	\end{subfigure}
    \centering
	\begin{subfigure}{0.225\linewidth}
		\centering
		\includegraphics[width=0.99\linewidth]{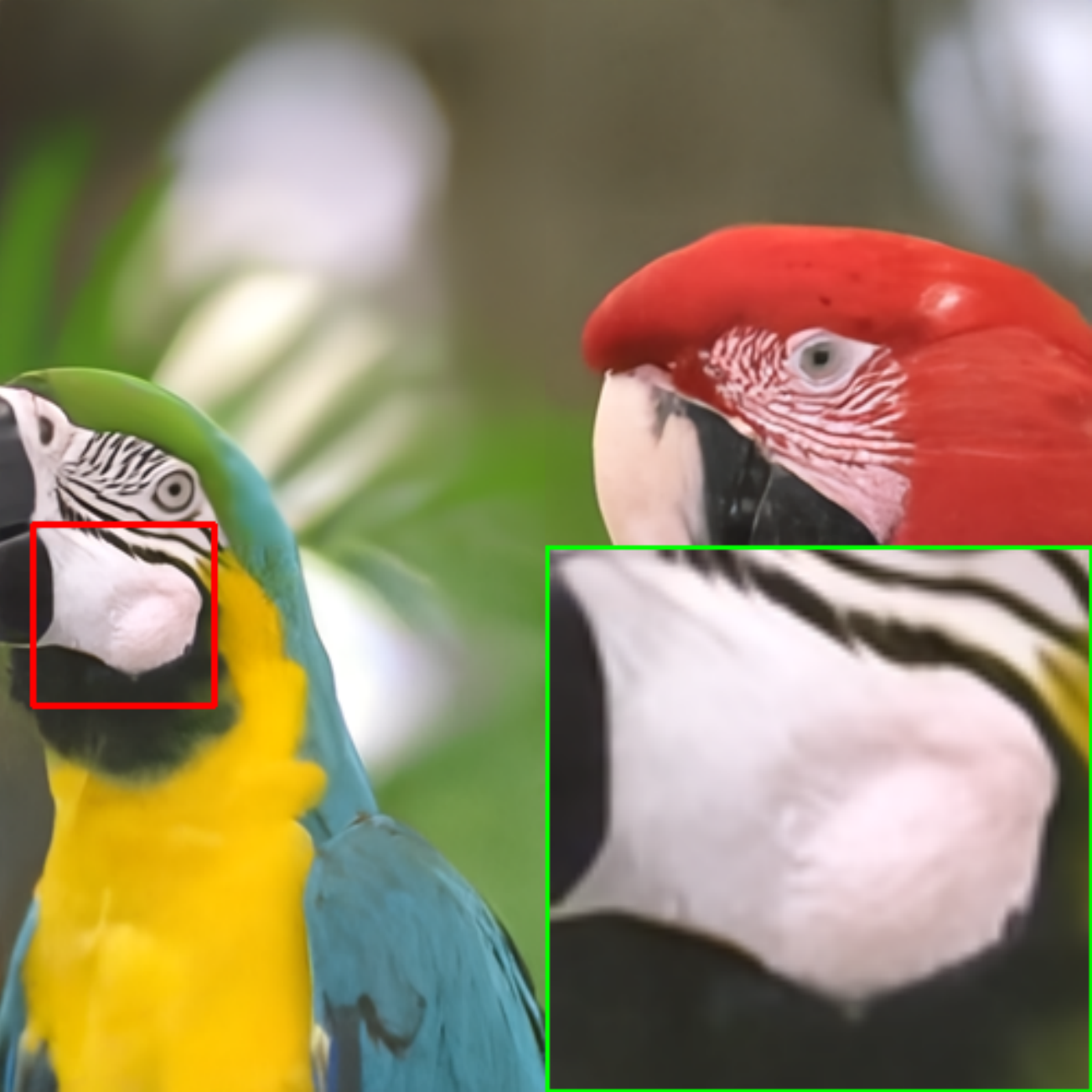}
		\caption{(\uppercase\expandafter{\romannumeral12})}
	\end{subfigure}
\caption{Visual results of synthetic noise removal for color image at the noise level 50. (\romannumeral1) Ground-truth image, (\romannumeral2) Noisy image / 14.16dB, (\romannumeral3) CBM3D / 30.95dB, (\romannumeral4) MCWNNM / 25.25dB, (\romannumeral5) CDnCNN-S / 31.47dB, (\romannumeral6) CDnCNN-B / 31.12dB, (\romannumeral7) IRCNN / 31.16dB, (\romannumeral8) BUIFD / 29.41dB, (\romannumeral9) FFDNet / 31.50dB, (\romannumeral10) ADNet / 31.47dB, (\romannumeral11) AirNet / 31.50dB, (\romannumeral12) DRANet / 32.03dB.}
\label{fig:kodim23}
\end{figure*}

The visual results of different models on the color image are presented in Fig. \ref{fig:kodim23}, where the ``kodim23'' image from the Kodak24 dataset was used for presentation, and the noise level is set to 50. One can find that MCWNNM generates poor denoising performance, and the CBM3D, CDnCNN-S, BUIFD, FFDNet, ADNet, and AirNet produce over-smooth visual results, and many image details are lost in their denoised results. The CDnCNN-B, IRCNN, and the proposed DRANet can reconstruct high-quality images and obtain a better balance between noise elimination and image detail retention. Moreover, the proposed DRANet can obtain a better PSNR result.

\subsection{Real noise removal evaluation}
The proposed DRANet is a blind image denoising model, we also evaluated the denoising performance of the DRANet model on real noisy images. The MCWNNM \cite{Xu2017}, TWSC \cite{Xu2018}, DnCNN-B \cite{Zhang2017}, CBDNet \cite{Guo2019}, RIDNet \cite{Anwar2019}, AINDNet \cite{Kim2020}, VDN \cite{Yue2019}, SADNet \cite{Chang2020}, DANet+ \cite{Yue2020}, DeamNet \cite{Ren2021}, VDIR \cite{Soh2022}, and CycleISP \cite{Zamir2020} were compared.

Table \ref{tab:SIDD_DND} reports the average PSNR and SSIM values of different denoising models on the SIDD validation set \cite{Abdelhamed2018} and the DND sRGB images \cite{Plotz2017}. Compared with other models, the DRANet model achieved very competitive performances on real noise removal.

\begin{table*}[htbp]
\centering
\caption{Quantitative results (PSNR and SSIM) of the real noise removal evaluation. The top two results are emphasized in red and blue, respectively.}
\label{tab:SIDD_DND}
\begin{tabular}{cccccccccccccc}
\hline
\multirow{2}*{Methods} & \multicolumn{2}{c}{SIDD} & \multicolumn{2}{c}{DND}\\
\cline{2-5}
 \multicolumn{1}{c}{}  & PSNR & SSIM & PSNR &  SSIM\\
\hline
MCWNNM \cite{Xu2017} & 33.40 & 0.879 & 37.38 & 0.929\\
\hline
TWSC \cite{Xu2018} & 35.33 & 0.933 & 37.94 & 0.940\\
\hline
DnCNN-B \cite{Zhang2017} & 23.66 & 0.583 & 37.90 & 0.943\\
\hline
CBDNet \cite{Guo2019} &  30.78 & 0.951 & 38.06 & 0.942\\
\hline
RIDNet \cite{Anwar2019} & 38.71 & 0.954 & 39.23 & 0.953\\
\hline
AINDNet \cite{Kim2020} & 38.95 & 0.952 & 39.37 & 0.951\\
\hline
VDN \cite{Yue2019} & 39.28 & \textcolor{blue}{0.956} & 39.38 & 0.952\\
\hline
SADNet \cite{Chang2020} & 39.46 & \textcolor{red}{0.957} & 39.59 & 0.952\\
\hline
DANet+ \cite{Yue2020} & 39.43 & \textcolor{blue}{0.956} & 39.58 & \textcolor{blue}{0.955}\\
\hline
DeamNet \cite{Ren2021} & 39.35 & 0.955 & \textcolor{blue}{39.63} & 0.953\\
\hline
VDIR \cite{Soh2022} & 39.26 & 0.955 & \textcolor{blue}{39.63} & 0.953\\
\hline
CycleISP \cite{Zamir2020} & \textcolor{red}{39.52} & \textcolor{red}{0.957} & 39.56 & \textcolor{red}{0.956}\\
\hline
DRANet & \textcolor{blue}{39.50} & \textcolor{red}{0.957} & \textcolor{red}{39.64} & 0.952\\
\hline
\end{tabular}
\end{table*}

\begin{figure*}[htbp]
	\centering
    \captionsetup[subfigure]{labelformat=empty}
	\begin{subfigure}{0.225\linewidth}
		\centering
		\includegraphics[width=0.99\linewidth]{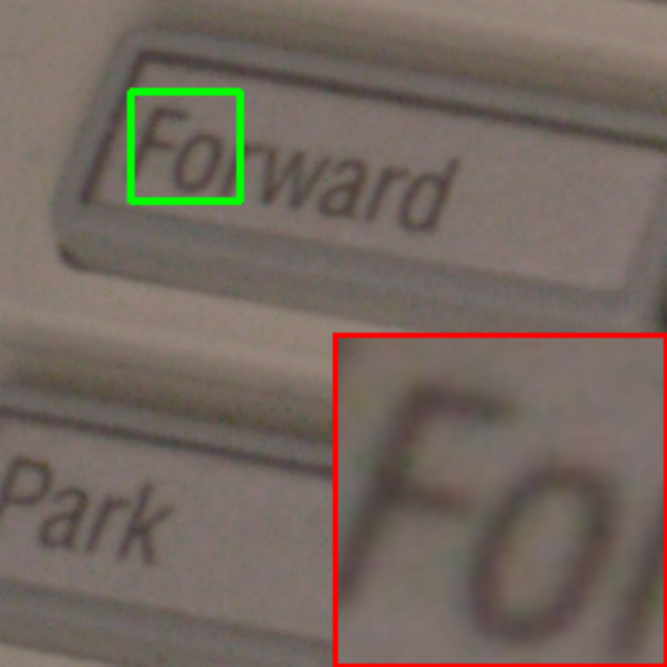}
		\caption{(\romannumeral1)}
	\end{subfigure}
    \centering
	\begin{subfigure}{0.225\linewidth}
		\centering
		\includegraphics[width=0.99\linewidth]{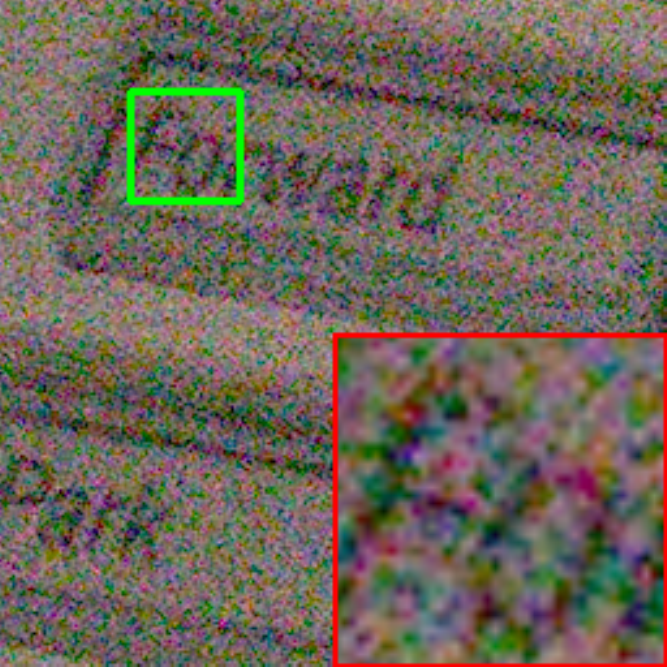}
		\caption{(\romannumeral2)}
	\end{subfigure}
    \centering
	\begin{subfigure}{0.225\linewidth}
		\centering
		\includegraphics[width=0.99\linewidth]{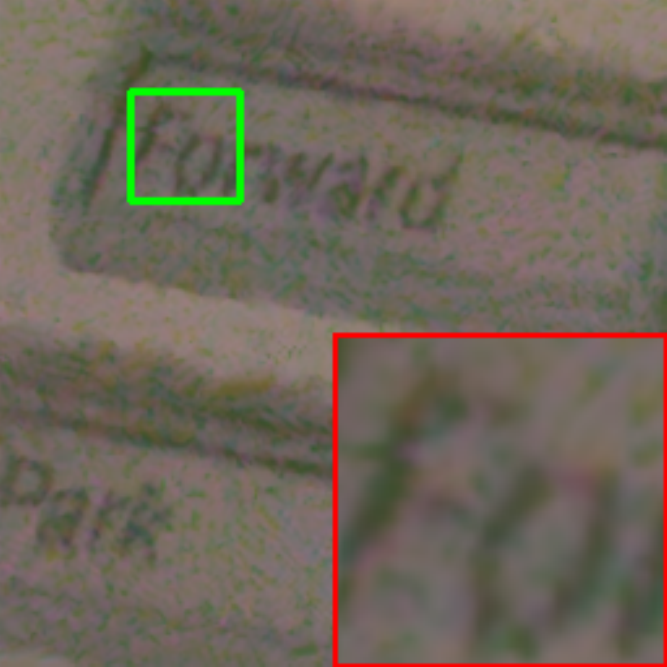}
		\caption{(\romannumeral3)}
	\end{subfigure}
    \centering
	\begin{subfigure}{0.225\linewidth}
		\centering
		\includegraphics[width=0.99\linewidth]{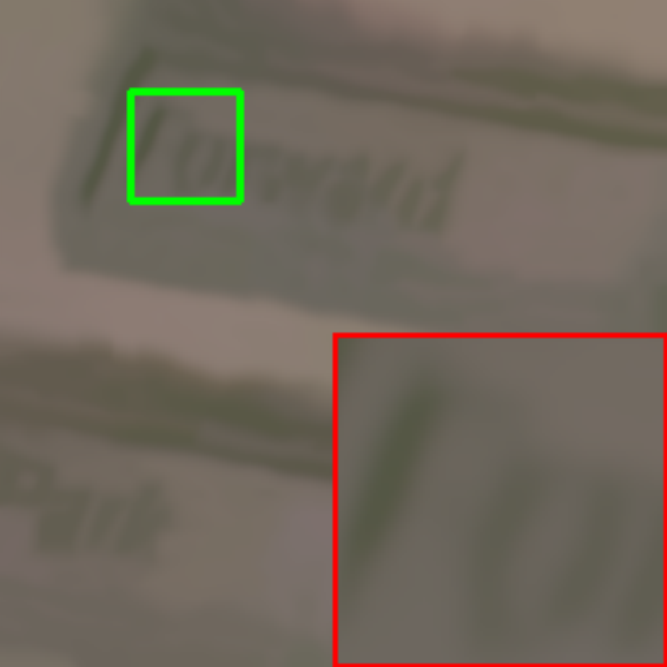}
		\caption{(\romannumeral4)}
	\end{subfigure}
    \centering
	\begin{subfigure}{0.225\linewidth}
		\centering
		\includegraphics[width=0.99\linewidth]{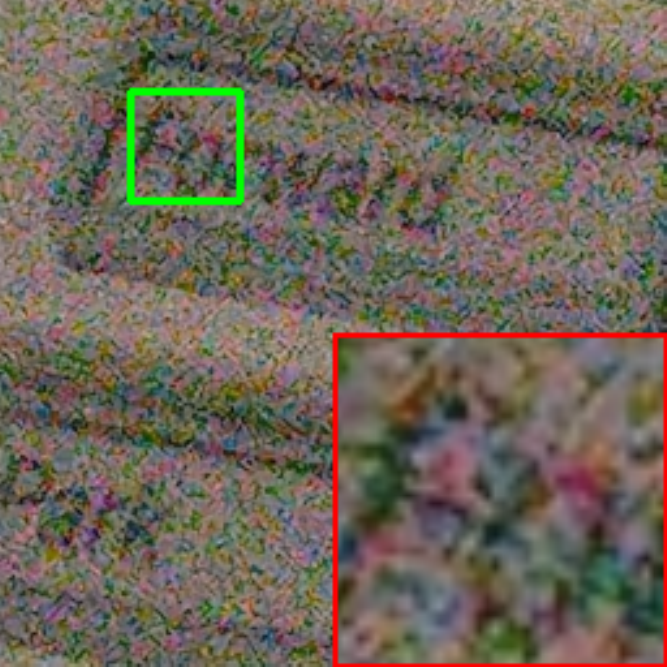}
		\caption{(\romannumeral5)}
	\end{subfigure}
    \centering
	\begin{subfigure}{0.225\linewidth}
		\centering
		\includegraphics[width=0.99\linewidth]{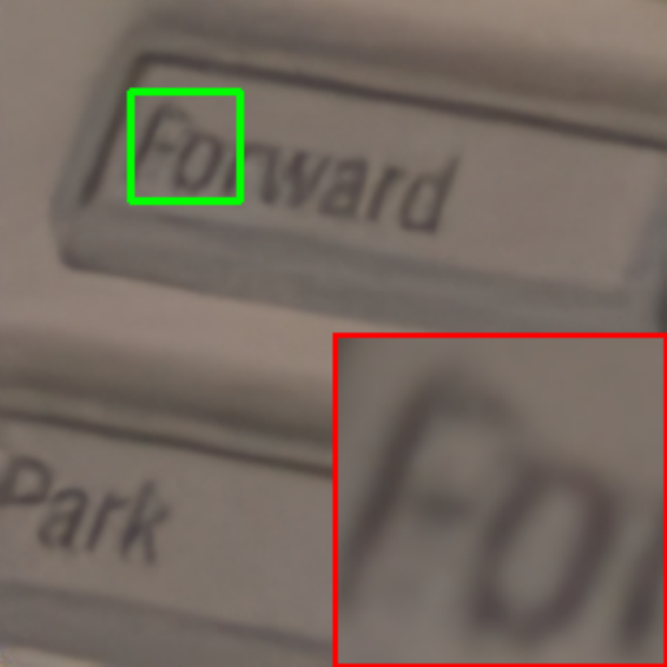}
		\caption{(\romannumeral6)}
	\end{subfigure}
    \centering
	\begin{subfigure}{0.225\linewidth}
		\centering
		\includegraphics[width=0.99\linewidth]{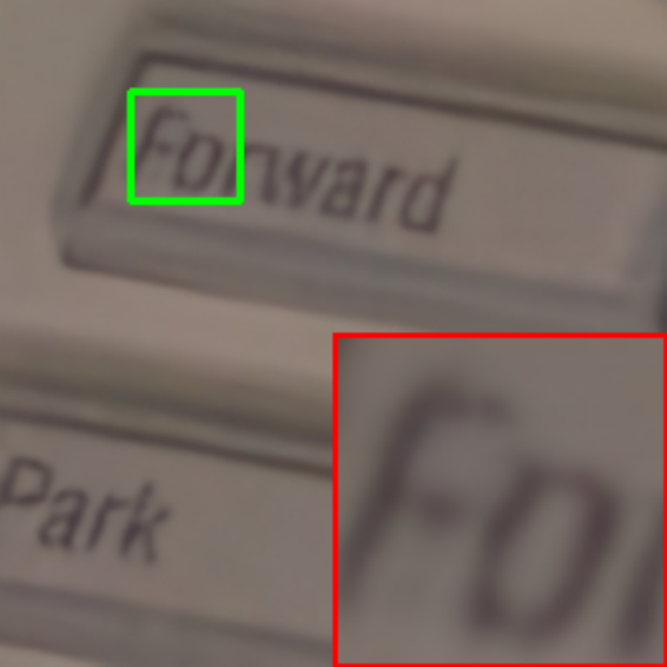}
		\caption{(\romannumeral7)}
	\end{subfigure}
    \centering
    \centering
	\begin{subfigure}{0.225\linewidth}
		\centering
		\includegraphics[width=0.99\linewidth]{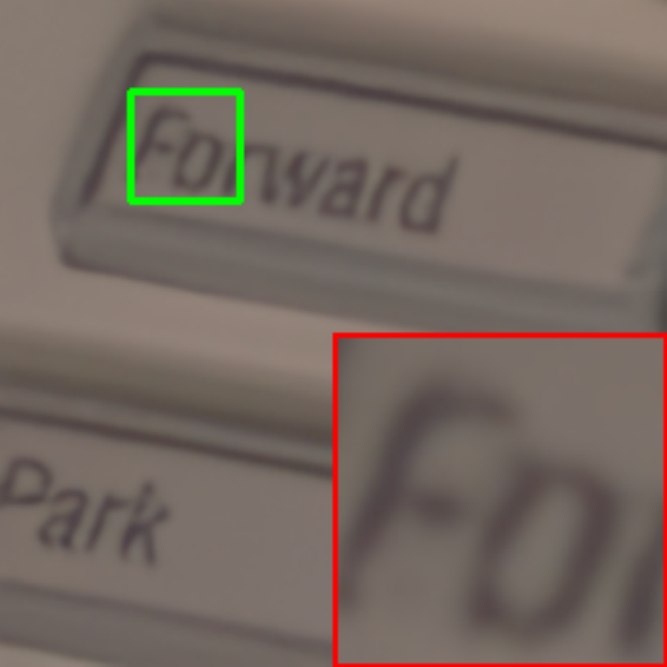}
		\caption{(\romannumeral8)}
	\end{subfigure}
    \centering
	\begin{subfigure}{0.225\linewidth}
		\centering
		\includegraphics[width=0.99\linewidth]{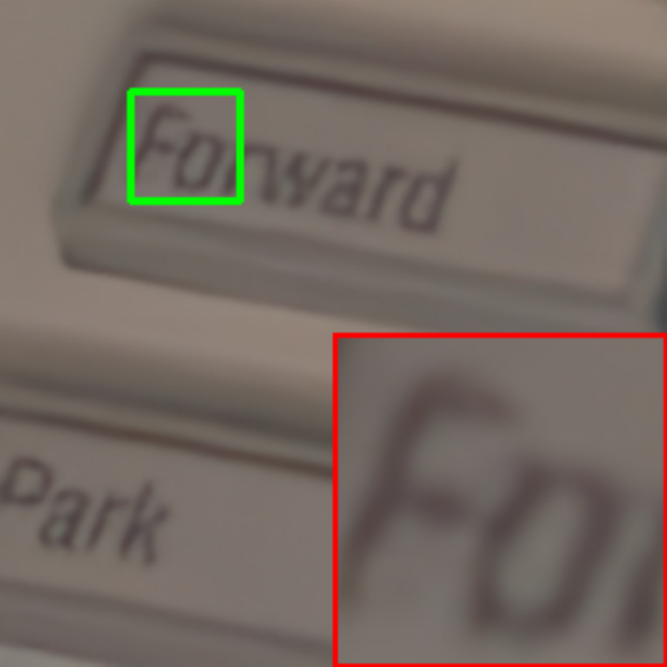}
		\caption{(\romannumeral9)}
	\end{subfigure}
    \centering
	\begin{subfigure}{0.225\linewidth}
		\centering
		\includegraphics[width=0.99\linewidth]{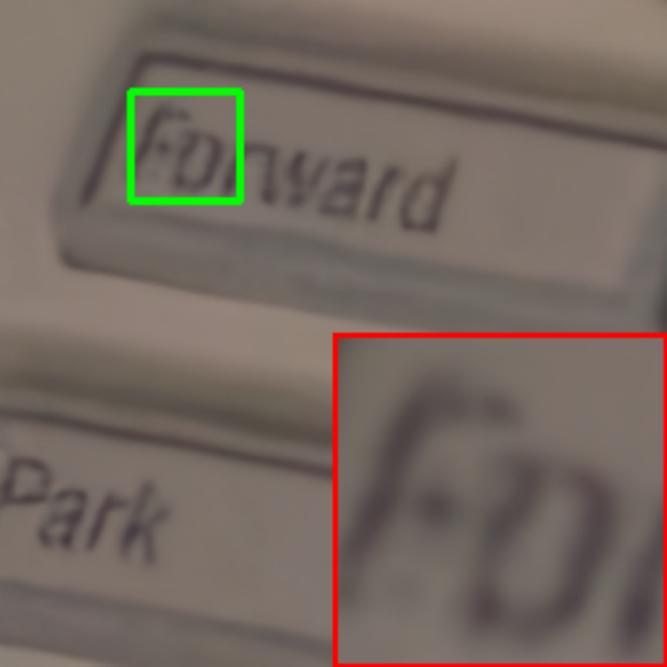}
		\caption{(\romannumeral10)}
	\end{subfigure}
    \centering
	\begin{subfigure}{0.225\linewidth}
		\centering
		\includegraphics[width=0.99\linewidth]{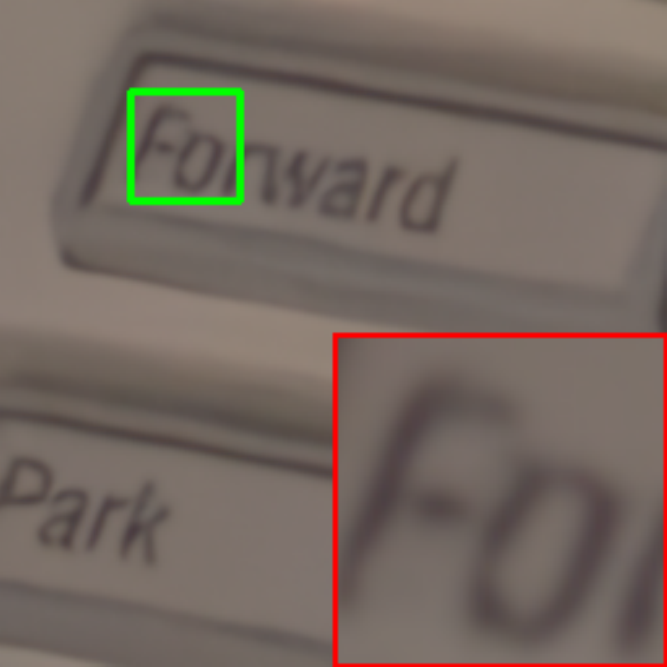}
		\caption{(\romannumeral11)}
	\end{subfigure}
    \centering
	\begin{subfigure}{0.225\linewidth}
		\centering
		\includegraphics[width=0.99\linewidth]{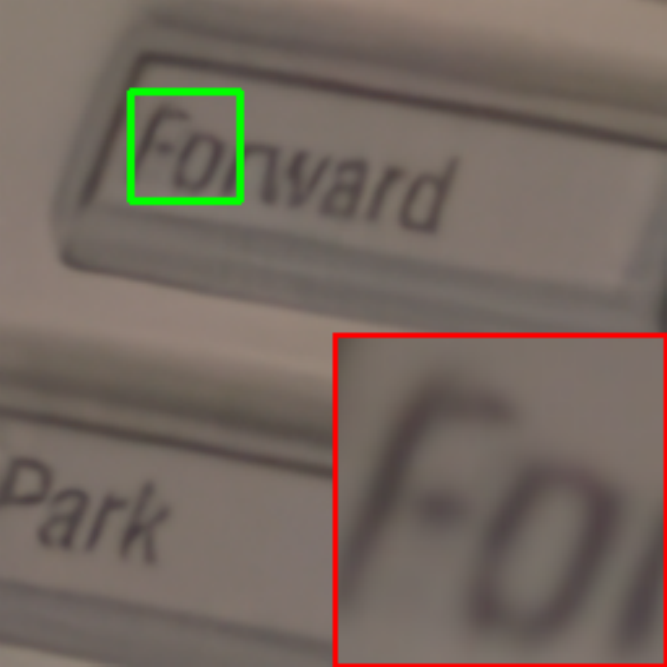}
		\caption{(\romannumeral12)}
	\end{subfigure}
\caption{Visual results of different denoising models on the real-world noisy image. (\romannumeral1) Ground-truth image / (PSNR/SSIM), (\romannumeral2) Noisy image / (18.25/0.169), (\romannumeral3) MCWNNM / (28.63/0.725), (\romannumeral4) TWSC / (30.42/0.837), (\romannumeral5) CDnCNN-B / (20.76/0.231), (\romannumeral6) AINDNet / (36.24/0.904), (\romannumeral7) VDN / (36.39/0.907), (\romannumeral8) SADNet / (36.72/0.910), (\romannumeral9) DANet+ / (36.37/0.908), (\romannumeral10) VDIR / (36.35/0.906), (\romannumeral11) CycleISP / (36.72/0.910), (\romannumeral12) DRANet / (36.72/0.917).}
\label{fig:10_3}
\end{figure*}

The visual effects of different models on real image denoising are displayed in Fig. \ref{fig:10_3}, where the ``10\_3'' image from the SIDD validation set was selected for comparison. One can see that MCWNNM and CDnCNN-B can hardly remove real noise, and the TWSC produces a blurred and over-smoothed result. In contrast, the AINDNet, VDN, SADNet, DANet+, VDIR, CycleISP, and the proposed DRANet achieve high-quality visual results. Moreover, the DRANet obtains the leading PSNR and SSIM values. It can be seen that for real-world noise removal, the DRANet also can produce very competitive results both qualitatively and quantitatively.

\subsection{Model complexity analysis}
In this subsection, we evaluated the model complexity by comparing model parameters and running time. The BM3D \cite{Dabov2007}, TNRD \cite{Chen2017}, MCWNNM \cite{Xu2017}, TWSC \cite{Xu2018}, DnCNN-B \cite{Zhang2017}, CBDNet \cite{Guo2019}, RIDNet \cite{Anwar2019}, AINDNet \cite{Kim2020}, VDN \cite{Yue2019}, SADNet \cite{Chang2020}, DANet+ \cite{Yue2020}, DeamNet \cite{Ren2021}, VDIR \cite{Soh2022}, CycleISP \cite{Zamir2020}, BUIFD \cite{Helou2020}, IRCNN \cite{ZhangZGZ2017}, FFDNet \cite{Zhang2018}, ADNet \cite{TianX2020}, BRDNet \cite{Tian2020}, DudeNet \cite{Tian2021}, ADNet \cite{TianX2020}, and AirNet \cite{Li2022} were utilized for comparison. The BM3D, TNRD, MCWNNM, and TWSC were implemented in the Matlab (R2020a) environment, other denoising models were implemented in the PyCharm (2021) environment.

The running time of the different models is displayed in Table \ref{tab:time}, where three synthetic images containing different channels with sizes $256\times256$, $512\times512$, and $1024\times1024$ were randomly chosen for evaluating the running speeds of different denoising models at noise level 25. The time values in Table \ref{tab:time} are the results obtained by calculating the average running time of 20 executions of each denoising model. It is worth noting that the memory transfer time between the CPU and GPU was ignored in our experiments.

One can find that compared with BM3D, TNRD, MCWNNM, AINDNet, VDN, AirNet, and VDIR, the DRANet yields competitive and faster denoising inference speed. In contrast, DRANet has a longer running time than DnCNN-B, BUIFD, IRCNN, FFDNet, ADNet, SADNet, DANet+, and CycleISP, however the DRANet can achieve better denoising performance than these models.

\begin{table*}[htbp]
\centering
\caption{Running time (in seconds) evaluation results on three synthetic noisy images.}
\label{tab:time}
\begin{tabular}{cccccccc}
\cline{1-8}
\multirow{2}*{Devices} & \multirow{2}*{Methods}	& \multicolumn{2}{c}{$256\times256$}	& \multicolumn{2}{c}{$512\times512$}	& \multicolumn{2}{c}{$1024\times1024$}\\
\cline{3-8}
 \multicolumn{1}{c}{} &  & Gray & Color & Gray & Color & Gray & Color\\
\cline{1-8}
\multirow{2}*{CPU} & BM3D \cite{Dabov2007} & 0.458	& 0.593	& 2.354	& 3.771	& 9.782	& 12.818\\
\cline{2-8}
    & MCWNNM \cite{Xu2017} & - & 62.777 & - & 277.623 & - & 1120.112  \\
\cline{1-8}
\multirow{14}*{GPU} & TNRD \cite{Chen2017}  & 0.843 & - & 1.418 & - & 3.589 & -\\
\cline{2-8}
    & DnCNN-B \cite{Zhang2017} & 0.032	& 0.032	& 0.037	& 0.037	& 0.057	& 0.057\\
\cline{2-8}
    & BUIFD \cite{Helou2020} & 0.035	& 0.037	& 0.050	& 0.053	& 0.112	& 0.123\\
\cline{2-8}
    & IRCNN \cite{ZhangZGZ2017}  & 0.030	& 0.030	& 0.030	& 0.030	& 0.030	& 0.030\\
\cline{2-8}
    & FFDNet \cite{Zhang2018} & 0.031	& 0.030	& 0.031	& 0.030	& 0.032	& 0.030\\
\cline{2-8}
    & ADNet \cite{TianX2020} & 0.031 & 0.033 & 0.035 & 0.045 & 0.051 & 0.093\\
\cline{2-8}
    & SADNet \cite{Chang2020}  & 0.030 & 0.030 & 0.043 & 0.044 & 0.101 & 0.102\\
\cline{2-8}
    & DANet+ \cite{Yue2020}  &  - & 0.025 & - & 0.027 & - & 0.041\\
\cline{2-8}
    & CycleISP \cite{Zamir2020}  &  - & 0.055 & - & 0.156 & - & 0.545\\
\cline{2-8}
    & AINDNet \cite{Kim2020} & - & 0.531 & - & 2.329 & - & 9.573\\
\cline{2-8}
    & VDN \cite{Yue2019} & 0.144 & 0.162 & 0.607 & 0.597 & 2.367 & 2.376\\
\cline{2-8}
    & AirNet \cite{Li2022}  &  - & 0.143 & - & 0.498 & - & 2.501\\
\cline{2-8}
    & VDIR\cite{Soh2022}  &  - & 0.385 & - & 1.622 & - & 6.690\\
\cline{2-8}
    & DRANet & 0.149 & 0.150 & 0.347 & 0.356 & 1.055 & 1.061\\
\cline{1-8}
\end{tabular}
\end{table*}

Table \ref{tab:Parameters} gives the numbers of model parameters of the compared models, where the number of parameters of each method is calculated by using the \textcolor{red}{pytorch-OpCounter package\footnote{\url{https://github.com/Lyken17/pytorch-OpCounter}}} for grayscale and color image, respectively. One can see that compared with many state-of-the-art image blind denoising models, such as the CBDNet, VDN, VDIR, DeamNet, AINDNet, AirNet, SADNet, DANet+, and CycleISP, the DRANet can obtain better denoising performances in most denoising tasks with a simpler model structure.

\begin{table*}[htbp]
\centering
\caption{The numbers of parameters (in K) of different denoising models.}
\label{tab:Parameters}
\begin{tabular}{ccc}
\hline
\multirow{2}*{Models} & \multicolumn{2}{c}{Parameters} \\
\cline{2-3}
 \multicolumn{1}{c}{}  & Gray & Color \\
\hline
TNRD \cite{Chen2017} & 27 & - \\
\hline
DnCNN-B \cite{Zhang2017} & 668 & 673 \\
\hline
BUIFD \cite{Helou2020} & 1075 & 1085 \\
\hline
IRCNN \cite{ZhangZGZ2017} & 186 & 188 \\
\hline
FFDNet \cite{Zhang2018} & 485 & 852 \\
\hline
ADNet \cite{TianX2020} & 519 & 521 \\
\hline
DudeNet \cite{Tian2021} & 1077 & 1079 \\
\hline
BRDNet \cite{Tian2020} & 1113 & 1117 \\
\hline
RIDNet \cite{Anwar2019} & 1497 & 1499 \\
\hline
CBDNet \cite{Guo2019} & - & 4365 \\
\hline
VDN \cite{Yue2019} & 7810 & 7817 \\
\hline
VDIR \cite{Soh2022}  & - & 2227 \\
\hline
DeamNet \cite{Ren2021} & 1873 & 1876 \\
\hline
AINDNet \cite{Kim2020} & - & 13764 \\
\hline
AirNet \cite{Li2022} & - & 8930 \\
\hline
SADNet \cite{Chang2020} & 3450 & 3451 \\
\hline
DANet+ \cite{Yue2020} & - & 9154 \\
\hline
CycleISP \cite{Zamir2020} & - & 2837 \\
\hline
DRANet & 1612 & 1617 \\
\hline
\end{tabular}
\end{table*}

\section{Conclusion and future work}\label{Conclusion}
In this paper, we propose a new dual residual attention network (DRANet) for image denoising. The DRANet model contains two parallel sub-networks with different structures to improve network learning ability. Moreover, the residual attention block and the hybrid residual attention block are designed and applied for the two sub-networks, respectively. Therefore, the DRANet can learn complementary image features, and can select appropriate features to improve model performance. Moreover, the long skip connections between the attention blocks, as well as the global features fusion between two sub-networks, are utilized to obtain the global image features better. The residual learning is utilized to accelerate network training and enhance the denoising performance of the network.

Compared with other state-of-the-art methods, DRANet achieves competitive denoising performance on different denoising tasks, and an attractive balance between network complexity and performance is also obtained, which makes the DRANet can be an option for practical image denoising tasks. In the near future, we will continue to explore the impact of different network structures on denoising performance, and the application of DRANet to other image processing tasks such as image deblurring and image super-resolution will also be investigated.

\section{Acknowledgements}
The work described in this paper is supported by the Natural Science Foundation of China under grant No. 61863037, No. 41971392, and the Applied Basic Research Foundation of Yunnan Province under grant No. 202001AT070077.
\bibliography{reference}
\end{document}